\documentclass[a4paper]{article}
\pdfoutput=1

\usepackage{standalone}
\usepackage{das}

\usepackage[numbers]{natbib}
\usepackage{amsfonts}
\usepackage{amsmath}
\usepackage{amssymb}
\usepackage{graphicx}
\usepackage{algorithm}
\usepackage{relsize}
\usepackage{siunitx}
\usepackage{upgreek}
\usepackage[hidelinks]{hyperref}
\usepackage{tikz}
\usetikzlibrary{automata}
\usetikzlibrary{matrix}
\usetikzlibrary{positioning}
\usetikzlibrary{shapes.geometric}
\usetikzlibrary{shapes.misc}
\usetikzlibrary{shapes.arrows}
\usetikzlibrary{shapes}     %?
\usetikzlibrary{fit}                    % fitting shapes to coordinates
\usetikzlibrary{backgrounds}    % drawing the background after the foreground
\usetikzlibrary{chains}
\usetikzlibrary{decorations.pathreplacing}

\let\emptyset\relax
\let\emptyset\varnothing

\usepackage{listings}
\usepackage{fancyvrb}

\definecolor{rltred}{rgb}{0.75,0,0}
\definecolor{dblue}{RGB}{31,73, 125}
\definecolor{mblue}{RGB}{79,129, 189}
\newcommand{\code}[1]{\lstinline[mathescape=true,basicstyle=\ttfamily]!#1!}

\newcommand{\mtxfig}{$\mu$-Tx}
\newcommand{\mtx}{$\upmu$-Tx}
\newcommand{\sa}{software agent}

\newcommand{\sm}{state machine}
\newcommand{\sms}{state machines}
\newcommand{\rbt}{Red-Black Tree}
\newcommand{\rbts}{Red-Black Trees}
\newcommand{\avlt}{AVL Tree}
\newcommand{\avlts}{AVL Trees}
\newcommand{\trees}{\rbts{} and \avlts{}}
\newcommand{\tree}{\rbt{} and \avlt{}}
\newcommand{\nvm}{NVRAM}
\newcommand{\redolog}{redo log}
\newcommand{\redologs}{redo logs}
\newcommand{\infiniband}{InfiniBand}
\DeclareMathAlphabet{\mathcal}{OMS}{cmsy}{m}{n} %% for \mathcal{P} correct symbol for failure detector

\begin{document}%

\title{Transactions on Red-black and AVL trees in \nvm{}}

\author{Thorsten Sch{\"u}tt \and Florian Schintke \and Jan Skrzypczak}

\date{%
    Zuse Institute Berlin\\%
%%    \today %% timestamp and date is given by arXiv
}

\maketitle

\begin{abstract}
% 4 Sentences:
% State the problem
Byte-addressable non-volatile memory (\nvm{}) supports persistent
storage with low latency and high bandwidth. Complex data structures
in it ought to be updated transactionally, so that they remain recoverable at
all times.
% Say why it's an interesting problem
Traditional database technologies such as keeping a separate log, a
journal, or shadow data work on a coarse-grained level, where the
whole transaction is made visible using a final atomic update
operation. These methods typically need significant additional space
overhead and induce non-trivial overhead for log pruning, state
maintenance, and resource (de-)allocation. Thus, they are not
necessarily the best choice for \nvm{}, which supports fine-grained,
byte-addressable access.

% Say what your solution achieves
We present a generic transaction mechanism to update dynamic complex
data structures `in-place' with a constant memory
overhead. It is independent of the size
of the data structure. We demonstrate and evaluate our approach on
\trees{} with a \redolog{} of constant size (4 resp. 2 cache lines).
% Say what follows from your solution
The \redolog{} guarantees that each accepted (started) transaction is
executed eventually despite arbitrary many system crashes and
recoveries in the meantime. We update complex data structures in local
and remote \nvm{} providing exactly once semantics and durable
linearizability for multi-reader single-writer access. To persist
data, we use the available processor instructions for \nvm{} in the
local case and remote direct memory access (RDMA) combined with a
\sa{} in the remote case.
\end{abstract}

%\keywords{Red-black trees, AVL trees, transactions, state machines, exactly once,
%%  linearizability}

\section{Introduction}

The introduction of \nvm{} enables a new range of applications, but it
also causes new challenges for their effective use. \nvm{} is the
first non-volatile storage providing byte-granular access with
low-latency and high bandwidth. In addition, it will replace SSDs as
fastest persistent storage in the storage hierarchy. While SSDs only
provide block-oriented APIs, \nvm{} comes as a standard DIMM. Plain
loads and stores suffice to get direct access (DAX) while bypassing
the operating system. For recoverability and consistency of data
structures, it becomes relevant when, in which order, and which part
of them will be written to \nvm{} from the processor's caches---either
explicitly by instructions or implicitly by cache evictions. This is
influenced by aspects such as data alignment, weak memory models, and
cache properties such as associativity, size, and its replacement and
eviction policy. To conquer all these aspects, complex data structures
stored in \nvm{} must be recoverable at all times, which requires new
and sound transactional update mechanisms.

The literature on \nvm{} has been mostly focusing on
B+trees~\cite{comer-b-trees79} with a high
radix~\cite{Chen:2015:PBT:2752939.2752947,210510} to minimize costs
for insert and remove. By using a high radix, expansive operations
like balancing happen seldom. Insert and remove are operations on
leaves. Thus, they do not have to be transactional and only require a
few persist calls.  If the sequence of calls is interrupted by a
crash, the tree remains valid. They are tuned for absolute performance.

In contrast, balancing is the common case for
\trees{}~\cite{TARJAN1983253}. Insert and remove always need an
unpredictable and variable sequence of operations, i.e., balancing,
recoloring, and updating the balance factors. This depends on the size
and the shape of the tree as they work across several levels of it.
For \nvm{}, the steps of the sequence have to happen atomically
despite an arbitrary number of crashes and restarts. Otherwise, the
tree might become invalid, unrecoverable, and may lose
sub-trees. Thus, transactions are needed~\cite{schutt2008scalaris}.

The literature so far focused on a copy-on-write style for updating
data-structures. It comes with additional costs for allocating,
de-allocating, and garbage collection. For B+ trees, a constant amount
of memory is needed, i.e., the size of a node, which simplifies the
process. As~\citet{Wang:2018:PRN:3190860.3177915} noted, for \rbts{}
copy-on-write operations touch almost the complete tree. One needs to
allocate and de-allocate a variable amount of memory for
operations. We split updates into a sequence of
micro-transactions. Thus a constant size \redolog{} suffices for all
operations. We neither allocate nor de-allocate memory for operations
as all updates are in-place~\cite{skrzypczak2020rmwpaxos}. It shows
its strengths for complex data structures where updates are global
operations and touch large parts of the data structure. There is no
doubt that performance-wise B+-trees beat binary trees. It is by
design. However, binary trees represent a wider class of dynamic data
structures using pointers. For binary trees, we needed to invent new
methodologies for storing data structures in \nvm{} that widely differ
from B+-trees. They could also be applied to data structures such as
(doubly) linked lists, priority queues, or graphs. Trees are often
used as proxies for index data
structures~\cite{schutt2008range,schutt2007structured}.

As \nvm{} behaves like memory rather than a spinning hard-disk, we can
use remote direct memory access (RDMA) of modern interconnects to
directly access \nvm{} on remote nodes. Local and remote access can
rely on a common set of operations. For \nvm{}, access can be
expressed in terms of read, write, atomic compare\&swap (CAS), and
persist operations. For remote access, get, put, remote atomic CAS,
and remote persist of the \emph{passive target communication} model
defined in the MPI standard~\cite{mpi31} can be used. For passive
target communication, the origin process can access the target's
memory without involvement of the target process. It is similar to a
shared memory model and allows the design of a single transaction
system based on common primitives for both local and remote \nvm{}.

We support exactly once
operations~\cite{DBLP:conf/ppopp/FriedmanHMP18} on dynamic complex
data structures in local and remote \nvm{} and make the following main
contributions:

\vspace{-\topsep}
\begin{itemize}
  \setlength{\itemsep}{0pt}
  \setlength{\parsep}{0pt}
\item We designed a new transaction system for \nvm{}, which splits
  large multi-step transactions into a sequence of
  micro-transactions. A \sm{} describes the transaction and the
  sequence of micro-transactions. Each micro-transaction resp.\ state
  transition is idempotent allowing atomicity and recovery
  in the failure case for multi-step transactions. All updates happen
  directly on the data structure `in-place' without shadow copying. It
  incrementally transforms the old data structure into the new
  one. All accepted operations will eventually succeed
  (\secref{sec.sm}).
\item A \emph{\redolog{}} of constant size (four resp.\ two cache
  lines for \trees{}) is used to guarantee recoverability and
  atomicity at all times. Note that the size of the \redolog{} is
  independent of the size of the data structure
  (\secref{sec.overhead}).
\item Our approach supports exactly once semantics and guarantees
  durable linearizability for all operations, both local and remote. Failed
  clients cannot corrupt any data (\secref{sec.durablelin}).
\item We implemented balanced \trees{} in \nvm{} using our
  approach---local and with passive target communication for remote
  access (\secref{sec.rbrdma}).
\item Intel guarantees 8~byte fail-safe atomicity for \nvm{}. For our
  approach 7~bytes suffice as we do not rely on atomic pointer updates
  (\secref{sec.sm}).
\item We designed a multi-reader single-writer lock with $f$-fairness
  to coordinate concurrent RDMA writes. Failed lock-holders can be
  safely expelled by other processes, because their process ids are stored
  in the lock, which allows other clients to use failure detectors
  (\secref{sec.locks}).
\item We simulated more than 2,000,000 power failures by killing
  processes to validate the robustness of our approach
  (\secref{sec.testing}).
\item Our evaluation shows more than 2,300/s key-value pair inserts
  into \rbts{} using passive target communication with \nvm{}. For
  \avlts{}, we reached more than 1,800/s inserts
  (\secref{sec.remote}). For local access, we reached almost 400,000 inserts
  per second (\secref{sec:evaluation:local}).
\end{itemize}

\section{System Model\label{sec.systemmodel}}

We assume a full-system failure
model~\cite{10.1007/978-3-662-53426-7_23}. On a crash, all transient
state (of all processes) is lost. Only operations on fundamental,
naturally aligned data types up to 8~bytes are atomic fail-safe in
\nvm{}, but 7~bytes are enough for our approach.

While RDMA operations can fail non-atomically, we assume 64~bit RDMA CAS
operations to be atomic. To detect failed nodes, we use the weak
failure detector $\Diamond\mathcal{W}$~\cite{chandra1996weakest}. We
consider a system with a single server storing data without
replication for simplicity. An arbitrary number of read/write clients
may try to access the data concurrently. We do not consider Byzantine
failures.

\section{Preliminaries}

As discussed above, hardware only supports atomic updates of
8~bytes. In the following, we describe the basic concepts needed for
larger updates and describe in detail the methods for persisting data
with \nvm{}.

\subsection{Logging and shadow copying\label{sec.shadow-copying}}

As long as updates are atomic, i.e., 8~bytes for \nvm{} or a block for
SSDs, they can be done in-place. For non-atomic updates, transaction
systems~\cite{Mohan:1992:ATR:128765.128770,Wang:2018:PRN:3190860.3177915,arulraj2015let}
use a combination of different techniques to preserve consistency in
the face of crashes. Logging uses undo and \redologs{} to store enough
data to roll-back an interrupted transaction (undo) or retry the
transaction again (redo). Undo logging tends to be more costly. It has
to log every store before executing it. Thus, redo logging is the
preferred technique. Some databases~\cite{arulraj2015let} use a
combination of both. Shadow copying, also known as copy-on-write,
creates a copy of the data to be updated, updates the copy, and
atomically replaces the old data with the new data. For example, to
atomically update a tree node, a copy of the node is created, updated,
and then the parent's pointer to the node is atomically
updated. Often, it is sufficient to replace one 8-byte pointer for the
last step, which can be done atomically.

\subsection{How to persist data with \nvm{}?\label{sec.howtopersist}}

Persisting data in \nvm{} works similar to block-oriented storage,
i.e., writes followed by a flush, but the details differ. Persisting
data to \nvm{} relies on cache line (cl) flushing, which persists data
as a side effect. According to~\citet{pcommit}, there are four
different methods: \code{clflush}, \code{movnti+sfence},
\code{clflushopt+sfence}, and \code{clwb+sfence}. The
\texttt{clflushopt} and \texttt{clwb} operations were recently
introduced with the Skylake micro architecture.  Stores before an
\code{sfence}, a store fence, are separated from following
stores. \texttt{clflush} and \texttt{clflushopt} persist the cache
line's data by invalidating it and writing the content back to
memory. \texttt{movnti} is a non-temporal store, which bypasses caches
and writes directly to memory. Hot data has to be loaded again from
memory. To mitigate the costs of cache misses, prefetching can be
used~\cite{nvtree}. A \code{clwb} ought to perform a write-back of the
data without invalidating the cache line.  It is recommended to use
\code{clwb}~\cite{pcommit}.

While any pair of \code{clflush} instructions are ordered, pairs of
\code{clflushopt} are only ordered with each other when accessing the
same cache line. Pairs of \code{clwb} always remain unordered with
each other. These recently introduced operations of the Skylake micro
architecture shown above allow more parallelism than the older ones.

Former versions of~\citet{pcommit3} recommended
\texttt{clflush+mfence}, where \texttt{mfence} is a memory fence that
separates load \emph{and} stores before and after the fence.
CDDS~\cite{Venkataraman:2011:CDD:1960475.1960480} and
NV-Tree~\cite{nvtree} even used \code{mfence+clflush+mfence}. Memory
fences have much higher latency than store fences.

\medskip

Today, there is no hardware support for remote persisting. The
standard work-around is proposed by SNIA~\cite{sniaha}. After an
RDMA write, the client has to request cache flushing by sending a
message to a \sa{}.  The \sa{} then must respond with an \emph{ack}
message. This workflow is implemented by the
PMDK.\footnote{http://pmem.io/pmdk/}

\section{Exactly once operations with Micro-Transactions (\mtx{}) and State
  Machines}\label{sec.sm}

Performing updates on complex data structures often requires a
sequence of smaller operations (recoloring, balancing, node splitting,
etc.). \nvm{} makes it challenging to perform them correctly in the
face of power losses as only aligned stores up to
8~bytes are fail-safe atomic with current hardware. All larger operations require
transactions. Otherwise, it is unknown which updates are persistent
in the failure case, i.e., reached the persistence domain. This can
fatally corrupt the data. Traditional techniques to address this problem
are shadowing, copy-on-write, and logging (see
\secref{sec.shadow-copying}).

\textbf{Aims.} Inserting or removing elements from trees, for
example, often requires a sequence of operations, such as tree
rotations (see \secref{sec.rbtrees}). We want to support such complex
data structure updates atomically `in-place'. We directly update the
data structure without shadow copying but with a \redolog{} of
\emph{constant size} independent of the size of the overall data
structure. The structure and actual size of such a constant-size
\redolog{} depends on the particular data structure and operations to
be supported. In \secref{sec.rbrdma}, we show some examples for \trees{}.

\textbf{Approach.} In general, we split an operation to be performed
on a complex data structure into a sequence of smaller operations,
which we execute in micro-transactions (\mtx{}) until the whole
operation is finished.  We want to be able to identify the ongoing
operation (insert or remove), detect the progress in that operation,
perform updates atomically, minimize the size of the \redolog{}, and
guarantee that all accepted operations will eventually complete. We
need the following components (see \figref{fig.rbtree-fsm}): (1) the
primary data structure \textsf{D} of potentially dynamic size
that we want to update, (2) a \redolog{} \textsf{L} of constant size,
and (3) a \sm{} $M$ with $\mathcal{S}$ states describing the sequence
of updates on \textsf{D} and \textsf{L}. \textsf{D} and \textsf{L} can
be seen as disjoint sets of byte ranges. \textsf{D}, \textsf{L}, and
the current state of $M$ are stored in \nvm{}.

For non-trivial updates, the idea is to establish a two-step mode of
operation repeatedly: First, persist all information that are required
to perform the operation on \textsf{D} in the redo log \textsf{L}. Afterwards,
perform the operation and persist it. Each step is idempotent until the next
micro-transaction begins. To separate
them from each other, a state variable is updated atomically
between steps to indicate which step finished last.

The current state of the \sm{} indicates whether the data structure is
clean for performing the next read or write or if it is currently performing
an ongoing concurrent operation. In this case, any upcoming read,
insert, or remove request has to wait. We reject concurrent operations
with locks (see \secref{sec.locks}) as our transaction approach cannot
handle concurrent accesses by itself. Otherwise, they may corrupt the data
structure~\cite{dechev2010understanding}.

\begin{figure}[h!]
  \centering
  \begin{tikzpicture}[>=stealth]
  \coordinate (origin) at (0,0);
  \coordinate (epochs) at (0,-2);

  \tiny

  \tikzstyle{clean}=[draw,circle, double,minimum size=4mm, inner sep=0pt]
  \tikzstyle{start}=[fill=gray!30!white,draw,circle, double,minimum size=4mm, inner sep=0pt]
  \tikzstyle{dirty}=[fill=gray!30!white,draw,circle, minimum size=4mm, inner sep=0pt]

  \node[clean] (clean0) {$c_0$};
  \node[above=0mm of clean0] {\emph{clean}};

  \node[dirty] (start0) [right=21mm of clean0] {$d_j$};
  \node[above=0mm of start0] {\emph{dirty}};
  \node[dirty] (dirty0) [right=23mm of start0]        {};
  \node[dirty] (dirty1) [right=22mm of dirty0]        {};
  \node[dirty] (dirty0b) [above=1mm of dirty0]        {};

  \draw[->,dashed] (clean0) to
  node[below] {\emph{1. \mtxfig{}}}
  node[below=.4,pos=-.15,anchor=north west,align=left] (ep1)
       {write $w_1$$\subseteq$\textsf{L}\\ persist(\,)}
  node[below=.4,pos=.55,anchor=north west,align=left] (ep2)
       {set$_{\mathit{atomic}}$ s\\persist(\,)~~~}
  (start0);

  \draw[->] (start0) --
  node[below] {\emph{2. \mtxfig{} (idempotent)}}
  node[below=.4,pos=-.05,anchor=north west,align=left] (ep3)
  {read $r_2$$\subseteq$\textsf{L}\\
    write $w_2$$\subseteq$\textsf{D}\\persist(\,)}
  node[below=.4,pos=.6,anchor=north west,align=left] (ep4)
    {set$_{\mathit{atomic}}$ s\\persist(\,)~~~}
  node[above,pos=.17,anchor=south] (operation)
    {\textsf{action}}
  node[above=.55,pos=.5,anchor=south,align=center] (controlflow)
    {\emph{decide next state by}\\\emph{if stmnt. or control flow}}
  coordinate[pos=.55] (fork)
  (dirty0);

  \draw[->,dashed] (dirty0) --
  node[below] {\emph{3. \mtxfig{} (idempotent)}}
  node[below=.4,pos=-.05,anchor=north west,align=left] (ep5)
  {read $r_3$$\subseteq$\textsf{L}$\cup$\textsf{D}~\\
    write $w_3$$\subseteq$\textsf{L}, \\~~$w_3\cap$$r_3$$=$$\emptyset$\\persist(\,)}
  node[below=.4,pos=.7,anchor=north west,align=left] (ep6)
       {set$_{\mathit{atomic}}$ s\\persist(\,)}
  coordinate[pos=.75] (join)
  (dirty1);

  \draw[->] (fork) to[out=0,in=180] (dirty0b);
  \draw[dashed] (dirty0b) to[out=0,in=180] (join);

  \draw[->,shorten >=1mm,bend right=20] (controlflow) to (fork);
  \draw[->,bend right,looseness=1.1] (dirty1) to
    node[above,sloped,pos=.1] {$\leftarrow$ read $r_4$$\subseteq$\textsf{L}}
    node[above,sloped,pos=.3] {$\leftarrow$ write $w_4$$\subseteq$\textsf{D}}
    node[above,sloped,pos=.5] {$\leftarrow$ persist(\,)}
    node[above=.5mm,sloped,pos=.75] {$\leftarrow$ set$_{\mathit{atomic}}$ s}
    node[above,sloped,pos=.9] {persist(\,)}
    node[above=3mm,sloped,pos=.5] {\emph{4. \mtxfig{} (idempotent)}}
  (clean0);

  \node[anchor=south] at (ep1 |- epochs) {Epoch 1};
  \node[anchor=south] at (ep2 |- epochs) {Epoch 2};
  \node[anchor=south] at (ep3 |- epochs) {Epoch 3};
  \node[anchor=south] at (ep4 |- epochs) {Epoch 4};
  \node[anchor=south] at (ep5 |- epochs) {Epoch 5};
  \node[anchor=south] at (ep6 |- epochs) {Epoch 6};

  \draw (ep1.north east) -- (ep1.north east |- epochs);
  \draw (ep2.north east) -- (ep2.north east |- epochs);
  \draw (ep3.north east) -- (ep3.north east |- epochs);
  \draw (ep4.north east) -- (ep4.north east |- epochs);
  \draw (ep5.north east) -- (ep5.north east |- epochs);

  \node[below=of ep1,anchor=north west] {persist(\,) $\in$ \{\code{clflush},
    \code{movnti+sfence}, \code{clflushopt+sfence}, \code{clwb+sfence}\}};

\end{tikzpicture}
  \caption{\mtx{} in a \sm{} with clean and dirty states working on a constant
    size \redolog{} \textsf{L}, a complex data structure \textsf{D} of arbitrary
    size, and the state variable s.\label{fig.state}}
\end{figure}
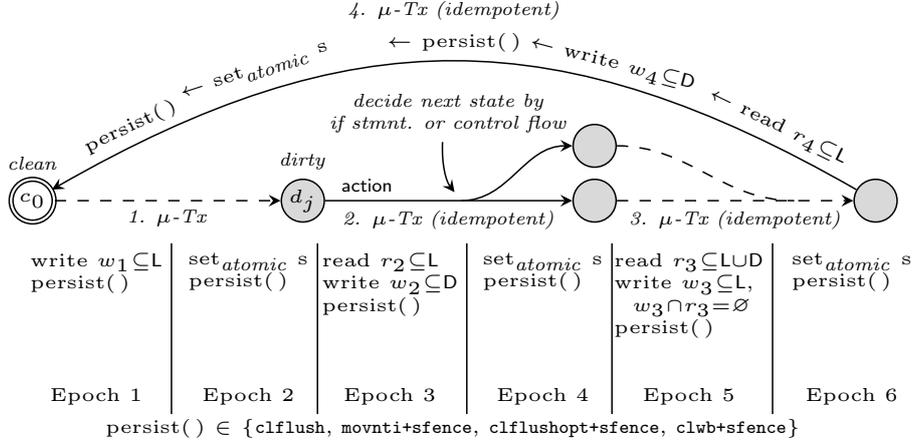

Each state transition of the \sm{} (see \figref{fig.state}) performs a
sequence of writes followed by a mechanism to make the writes
persistent---an epoch~\cite{Condit:2009:BIT:1629575.1629589}. A state
transition either updates \textsf{D} or \textsf{L} and then updates
the current state atomically. Thus, it consists of two epochs. We
require all state transitions to be idempotent. If a transition updates
\textsf{D}, there must be enough information in \textsf{L} to be able
to redo the operation. If it updates \textsf{L}, there must be enough
information in \textsf{D} and \textsf{L} to redo the operation, i.e.,
\textsf{D} can be the \redolog{} for \textsf{L}. Switching the roles
of the two components during an ongoing transaction is uncommon for
traditional databases, but it is one of the fundamental concepts of
our approach. Idempotence of state transitions facilitates applying
non fail-safe atomic updates consistently.

\textbf{State transitions.} A transition from the current state $s_i$
to a direct successor state $s_j$ is split into two steps: (a) an
epoch updating \textsf{D} or \textsf{L} and (b) atomically making $s_j$ the
current state. We discuss in \secref{sec.clocks} our
approach for identifying and atomically changing the current state. If
a client succeeds in updating \textsf{D} or \textsf{L}, it tries to
make $s_j$ the current state. When the client fails to complete the
first step, $s_i$ remains the current state. If the client finishes
the epoch and fails to make $s_j$ the current state, $s_i$ also
remains the current state. So, any attempt to perform a state
transition may fail. As all update operations resp.\ epochs are
required to be idempotent, a state transition can be blindly retried
arbitrarily often. Given enough progress, it will eventually
succeed. It will never abort.

To be able to precisely track progress and detect failures, we divide
the states $\mathcal{S}$ into disjoint sets of clean states
$\mathcal{C}$ and dirty states $\mathcal{D}$, such that
$\mathcal{S}=\mathcal{C}\cup \mathcal{D}$ and $\mathcal{C}\cap
\mathcal{D}=\emptyset{}$. If the \sm{} is in a clean state, clients
can read \textsf{D} or start a new transaction. If it is in a dirty
state, a transaction is ongoing and it has to be finished first.

\textbf{Recovery.} Recovery is agnostic to the ongoing operation. In
case of a power failure, we can always identify the current state at
all times. If we recover in the clean state, there was either no
accepted operation or an operation completed that did not return to
the client before the power loss. If we recover in a dirty state, we
continue the current operation until we reach the clean state as all
state transitions are idempotent.

For databases, recovery is separated from production, see
\secref{sec.relatedwork}. For recovery, databases have to analyze the
logs and cleanup the damage caused by the crash. In our approach, we
simply continue the ongoing operation. Recovery barely differs from
production.

In contrast to shadowing and logging, the \sm{} approach is less
affected by failures. When it reached the first dirty state, it can
guarantee the client that the operation will eventually succeed. The
first state transition accepts the operation and the following ones
perform the operation. Shadowing can only guarantee success after
completion. We can identify the kind of operation uniquely by the
current state. Different operations will use disjoint sets of dirty
states. If the machine is in a clean state, there is no ongoing
operation.

The recovery cost is negligible. We might lose one epoch, which we
have to repeat. The next process can continue where the last process
crashed. For comparison, NV-Trees~\cite{nvtree} only store the leaves
in \nvm{}. On recovery, they have to recompute the inner nodes.

\subsection{Overhead\label{sec.overhead}} There is a trade-off
between the number of states and the size of the \redolog{}. Larger
\sms{} tend to have smaller logs. Less information is needed to make
state transitions idempotent. However, they require more state
transitions and flushes. Smaller \sms{} will need larger logs and
require fewer state transitions and flushes.

The smallest \sm{} with one clean and one dirty state akin to
traditional transactions in databases has no constant size
\redolog{}. Remove for \avlts{} has in the worst case
$\mathcal{O}(\log{}n)$ balancing operations, which cannot be executed
in-place. All balancing operations require logging, see
\secref{sec.rotations}. Thus, the minimal \sm{} with two states has no
constant size \redolog{}. For remove with \avlts{}, the log's size is
a function of the depth of the tree. In our approach, it suffices to
provide enough space in the \redolog{} to make state transitions
idempotent. The complexity of state transitions is independent of the
size of the data structure.

There are operations that imperative code implements in-place, such as
tree rotations, updating pointers, and updating loop counters. These
operations are not idempotent and have to be split into two state
transitions. The smallest \sm{} with a constant size \redolog{}
depends on the complexity of the algorithm. Note that both \sms{} for
\avlts{} are larger than the two for \rbts{}. \avlts{} are
stricter balanced . Furthermore, the remove \sm{} for \avlts{} is
larger than the one for insert (35 vs. 24 states).

Despite all state transitions being idempotent, there are two
strategies for re-executing an interrupted state transition: (a)
blindly re-executing it and (b) analyzing the progress made before the
crash and only executing missing parts. In our implementation, we
used the former strategy. The number of stores per state
transition is small. A single rotation needs two state transitions
with three stores each. The gain of the latter strategy would be
negligible. For store intensive state transitions, like, e.g., writing
a GiB of data, the latter would be more efficient than the former.

\subsection{Maintaining the Current State\label{sec.clocks}}

State change operations have to be either fail-safe atomic or use a
\sm{}, otherwise we can reach unintended resp.\ unreachable states.

We use a global state variable of 7~bytes that stores the current
state and is updated atomically with CAS operations. This means that
the number of states is limited to $2^{56}$. When $2^{56}$ states are
insufficient, we create a \sm{} where \emph{D} is a variable with more
than 56~bits. However, the number of executable state transitions is
not bounded by the 7~byte limit. Unfortunately, this approach does not
allow reliably detecting whether the \sm{} made progress. Even if the
state variable did not change between two reads, the \sm{} may have
transitioned through several states in the meantime ending in the
original state again. Supporting detection of progress would require
an approach based on arbitrary precision counters, which is
unfortunately challenging with \nvm{}.

\subsection{Detectable Execution\label{sec.detectable}}

The \emph{detectable execution} property was introduced by
\citet{DBLP:conf/ppopp/FriedmanHMP18}. A data structure with this
property can, upon recovery after a crash, detect for each operation
executed during the crash whether it completed or aborted. This
property is necessary to execute operations exactly once.

To fulfill this property, additional state provided by the client is
required. The operation has to go through three steps: (1) announced,
(2) accepted, and (3) done. The client has to durably store its intent
to execute an operation (step 1). Then it can move the \sm{} from the
clean state into the first dirty state (the operation is accepted) and
durably store that the operation is accepted (step 2). Finally, the
client traverses the \sm{} until it reaches the clean state again and
durably marks the operation as done (step 3).

If a crash happens before step 1, the operation was never intended to
be executed. If the crash happens between step 1 and step 2, the \sm{}
is either in the clean state or the first dirty state. Thus, the
operation did not happen yet. If the crash happens between step 2 and
step 3, the \sm{} is either in a dirty state or the clean state. When
it is in a dirty state, the operation did not finish \fs{ok, so our
  lock to prevent concurrency then must be outside of this three steps
  state machine, otherwise the \sm{} could have started the next
  operation for another client already}. If it is in the clean state,
the operation finished, but the client did not update the state. If
the \sm{} is after step 3, the operation finished. Thus, we can
support exactly once semantic. We did not implement this approach and
the evaluation in \secref{sec:evaluation} does not consider it.

For systems based on shadow copying, it is challenging to support this
property. Updates become only visible with the last atomic pointer
update. Intermediate steps cannot be detected. Even if the client
durably stored its intent to perform an operation, there are no
visible events that allow distinguishing between success and failure
after the crash. For a no-op operation, we cannot detect whether it
succeeded or not. The updated value does not change and we cannot use
the value change to distinguish between success or failure.

\subsection{Durable Linearizability\label{sec.durablelin}}

According to \emph{durable linearizability} introduced by
\citet{10.1007/978-3-662-53426-7_23}, operations have to become
durable before they return. In case of a crash, all previously
completed operations remain completed. Operations started before the
crash may be visible, because they progressed enough but did not
return. The concept preserves important properties from
linearizability, composability, and non-blocking
progress~\cite{Herlihy:1990:LCC:78969.78972,DBLP:journals/debu/AguileraT16,skrzypczak2019linearizable}.

In contrast, \emph{buffered durable
  linearizability}~\cite{10.1007/978-3-662-53426-7_23} only requires
operations to be persistently ordered before they return. After a
crash, the data would still be persistent but not necessarily up to
date.

In our approach, all operations start at the state transition the last
operation finished. They start from the clean state, traverse the
\sm{}, and return to the clean state before they return. The current
state of \textsf{D} represents the execution resp.\ history of all
previously completed \mtx{}. If the \sm{} is in a dirty state after
the crash, an operation started but did not return. Given enough
progress, the \sm{} might be in the clean state after the crash, but
it did not return before the crash. Thus, our approach supports
durable linearizability.

\section{Self-balancing Binary Search Trees\label{sec.trees}}

Binary search trees are a widely studied topic. They serve as a
research vehicle for theoretical computer science and have many
applications in practical systems.

\subsection{\rbts{}\label{sec.rbtrees}}

\rbts{}~\cite{redblacktrees,cormen2009introduction} are half-balanced
binary trees. They have the following properties: (a) each node is
colored either red or black, (b) all children of red nodes are black,
and (c) every path from a given node to its leaves has the same number
of black nodes.  Violating property (b) or (c) is called a \emph{red
  violation} or \emph{black violation}, respectively. The shortest
possible path from a given node to its leave has all black nodes. The
longest possible path from a given node to its leave alternates
between black and red nodes with twice the length of the shortest
possible path. Thus, \rbts{} are called
half-balanced~\cite{olivie1982new}. Our root is always colored black
to prevent red violations (property (b)) between the root and its
children. Insert, remove, and lookup are all in
$\mathcal{O}(\log{n})$.

We focus on insert and remove as lookups are not of particular
interest regarding failures. We use top-down algorithms, which start
at the root and walk down the tree. They will always insert (add) or
remove a leaf object. On the way down, they anticipate and repair red
and black violations. This relies on node recoloring and tree
rotations. In case of a failure, recoloring is trivial to repair---we
verify the \rbt{} properties and recolor the nodes as
necessary. Failures during tree rotations are more challenging,
because they might fatally damage the tree and risk losing
sub-trees. Thus, tree rotations have to be performed atomically (see
\secref{sec.rotations}).

\subsection{\avlts{}\label{sec.avlrees}}

\avlts{}~\cite{avltrees} are balanced binary trees. In contrast to
\rbts{}, they are height balanced. For each node, the heights of the
subtrees of its two children differ at most by one. \rbts{} store a
color in each node to support the balancing scheme. \avlts{} store two
bits in each node---the balance. It is either -1, 0, or 1 depending on
the comparison of the height of the subtrees. \avlts{} also rely on
tree rotations for balancing. Insert, remove, and lookup are all in
$\mathcal{O}(\log{n})$.

\citet{Knuth:1998:ACP:280635} describes a top-down algorithm for
insert with $\mathcal{O}(1)$ single or double
rotations on average~\cite{mehlhorn1986amortized}. It walks down the tree and
inserts the new node at the bottom. On the way down, it records the
positions, which have to potentially be rebalanced. Additionally, it
adapts the respective balance factors.

Remove deviates from the algorithms described so far. It also requires
$\mathcal{O}(1)$ balance operations on average, but it can need up to
$\mathcal{O}(\log{n})$ balance
operations~\cite{tsakalidis1985rebalancing}. It must walk back up to
balance the tree. B and B+ trees~\cite{bayer1972symmetric,
  comer-b-trees79,Bayer1972} use a similar concept. If a
node is full, they split the node and walk up the tree.

The imperative C code from Julienne
Walker~\footnote{http://eternallyconfuzzled.com/tuts/datastructures/jsw\_tut\_rbtree.aspx.
}
uses a stack of size $\mathcal{O}(\log{n})$ for remove. This violates
our aim that the \redolog{} is constant size. On the way down, we
store pointers to parents in the nodes, which can be used to walk back
up the tree as needed. In general, it shows a limit of our
approach. Algorithms that need auxiliary space larger than
$\mathcal{O}(1)$ in the log cannot be directly supported when the log
must remain of constant size. Here, we used the common technique of
storing pointers to parents in the nodes. We discuss this problem in
more in detail in \secref{sec.limits}.

\subsection{Weight-balanced trees}

\trees{} use the height of sub-trees as a balancing
criterium. Weight-balanced
trees~\cite{doi:10.1137/0202005,hirai_yamamoto_2011} store the number
of nodes in its sub-trees in each node. A node $n$ is
$\alpha$-weight-balanced if $\text{weight}(\text{n.left}) \geq \alpha
* \text{weight}(n) \land \text{weight}(\text{n.right}) \geq \alpha *
\text{weight}(n)$. A larger $\alpha$ makes a tree more
balanced. Weight-balanced trees also rely on tree rotations for balancing.

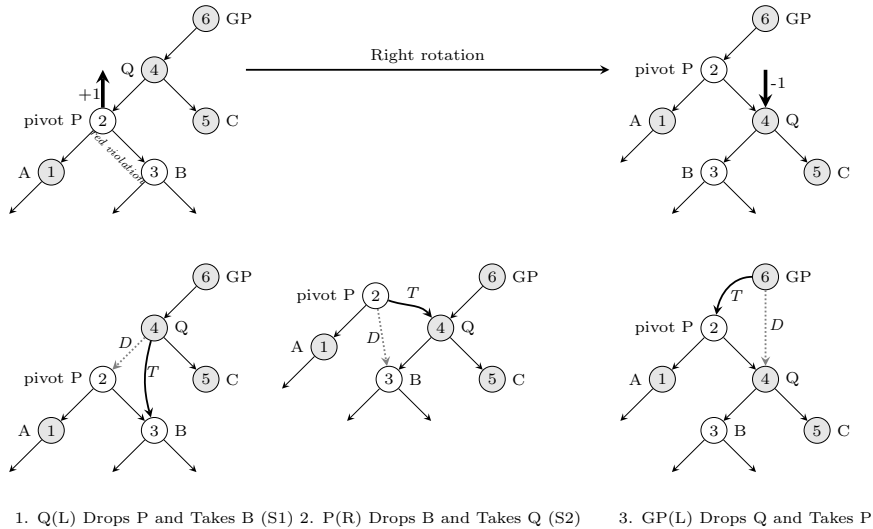
\begin{figure}[t!]
  \centering
  \begin{tikzpicture}[>=stealth]
  \coordinate (origin) at (0,0);
  \scriptsize

  \begin{scope}[start chain=p going right]
    \node [node distance=0mm,on chain=p] (alignment) {};
    \node [node distance=0mm,on chain=p] (left) {};
    \node [node distance=42mm,on chain=p] (middle) {};
    \node [node distance=40mm,on chain=p] (right) {};
  \end{scope}

  %% protocol
  \begin{scope}[
      start chain=alignment going below,
      start chain=left going below,
      start chain=middle going below,
      start chain=right going below,
      node distance=2mm]

    %% start time lines
    \foreach \x in {alignment, left, middle, right}
      \chainin (\x) [on chain=\x];

    \tikzstyle{worker}=[draw,circle,minimum size=4mm, inner sep=0pt]
    \tikzstyle{slave}=[fill=gray!20,draw,circle,minimum size=4mm, inner sep=0pt]

    %% alignment
    \node[         node distance = 80mm, on chain=alignment] (alignment-end) {};

    %% roots
    \node[         node distance = 0mm, on chain=left]  (left-root) {};
    \node[         node distance = 0mm, on chain=middle] (middle-root) {};
    \node[         node distance = 0mm, on chain=right] (right-root) {};

    %% left tree
    %% \node[slave] (lr) at (left-root) {GP};
    \node[slave] (lr1) [below left  = 0.0cm and 0.0cm of left-root, label=right:GP] {6};
    %% [below left  = 0.5cm and 0.5cm of left-root, label=left:Q]
    \node[slave] (ql1) [below left  = 0.5cm and 0.5cm of lr1, label=left:Q]  {4};
    \node[worker] (pl1) [below left  = 0.5cm and 0.5cm of ql1, label=left:pivot P]        {2};
    \node[slave] (al1) [below left  = 0.5cm and 0.5cm of pl1, label=left:A]         {1};
    \node[worker] (bl1) [below right  = 0.5cm and 0.5cm of pl1, label=right:B]       {3};
    \node[slave] (cl1) [below right  = 0.5cm and  0.5cm of ql1, label=right:C]       {5};

    \node[] (lal1) [below left  = 0.5cm and  0.5cm of al1]        {};
    \node[] (lbl1) [below left  = 0.5cm and  0.5cm of bl1]        {};
    \node[] (rbl1) [below right  = 0.5cm and  0.5cm of bl1]        {};

    \draw[->] (lr1) -- (ql1);
    \draw[->] (ql1) -- (pl1);
    \draw[->] (pl1) -- (al1);
    \draw[->] (pl1) -- node[below=1.6pt, sloped] {\smaller\smaller\emph{red violation}} (bl1);
    \draw[->] (ql1) -- (cl1);

    \draw[->] (al1) -- (lal1);
    \draw[->] (bl1) -- (lbl1);
    \draw[->] (bl1) -- (rbl1);

    %% bottom left tree
    %% \node[slave] (lr) at (left-root) {GP};
    \node[slave] (mr2) [below left  = 4.0cm and 0.0cm of left-root, label=right:GP] {6};
    %% [below left  = 0.5cm and 0.5cm of left-root, label=left:Q]
    \node[slave] (qm2) [below left  = 0.5cm and 0.5cm of mr2, label=right:Q]  {4};
    \node[worker] (pm2) [below left  = 0.5cm and 0.5cm of qm2, label=left:pivot P]        {2};
    \node[slave] (am2) [below left  = 0.5cm and 0.5cm of pm2, label=left:A]         {1};
    \node[worker] (bm2) [below right  = 0.5cm and 0.5cm of pm2, label=right:B]       {3};
    \node[slave] (cm2) [below right  = 0.5cm and  0.5cm of qm2, label=right:C]       {5};

    \node[] (lam2) [below left  = 0.5cm and  0.5cm of am2]        {};
    \node[] (lbm2) [below left  = 0.5cm and  0.5cm of bm2]        {};
    \node[] (rbm2) [below right  = 0.5cm and  0.5cm of bm2]        {};

    \draw[->] (mr2) -- (qm2);
    %%\draw[thick,->] (mr) to[out=-180,in=80] node [pos=0.1,below=-1pt] {\emph{S1}} (pm);
    \draw[thick,->,gray,densely dotted] (qm2) -- node [black,pos=0.6,above=1pt] {\emph{D}} (pm2);
    \draw[thick,->] (qm2) to[out=-105,in=105] node[pos=0.4,xshift=3pt] {\emph{T}} (bm2);
    \draw[->] (pm2) -- (am2);
    \draw[->] (pm2) -- (bm2);
    %%\draw[thick,->] (pm) to[out=-0,in=-75] node [pos=0.15,below=-1pt] {\emph{S2}} (qm);
    \draw[->] (qm2) -- (cm2);

    \draw[->] (am2) -- (lam2);
    \draw[->] (bm2) -- (lbm2);
    \draw[->] (bm2) -- (rbm2);

    %% bottom middle  tree
    %% \node[slave] (lr) at (left-root) {GP};
    \node[slave] (mr3) [below left  = 4.0cm and 0.0cm of middle-root, label=right:GP] {6};
    %% [below left  = 0.5cm and 0.5cm of left-root, label=left:Q]
    \node[slave] (qm3) [below left  = 0.5cm and 0.5cm of mr3, label=right:Q]  {4};
    \node[worker] (pm3) [below left  = 0.0cm and 1.5cm of mr3, label=left:pivot P]        {2};
    \node[slave] (am3) [below left  = 0.5cm and 0.5cm of pm3, label=left:A]         {1};
    \node[worker] (bm3) [below left  = 0.5cm and 0.5cm of qm3, label=right:B]       {3};
    \node[slave] (cm3) [below right  = 0.5cm and  0.5cm of qm3, label=right:C]       {5};

    \node[] (lam3) [below left  = 0.5cm and  0.5cm of am3]        {};
    \node[] (lbm3) [below left  = 0.5cm and  0.5cm of bm3]        {};
    \node[] (rbm3) [below right  = 0.5cm and  0.5cm of bm3]        {};

    \draw[->] (mr3) -- (qm3);
    %%\draw[thick,->] (mr) to[out=-180,in=80] node [pos=0.1,below=-1pt] {\emph{S1}} (pm);
    %%\draw[->] (qm) --  (pm);
    \draw[->] (qm3) -- (bm3);
    \draw[->] (pm3) -- (am3);
    \draw[thick,->,gray,densely dotted] (pm3) -- node[black,pos=0.5,xshift=-4pt] {\emph{D}} (bm3);
    \draw[thick,->] (pm3) to[out=-20,in=130] node [pos=0.5,above=0.5pt] {\emph{T}} (qm3);
    \draw[->] (qm3) -- (cm3);

    \draw[->] (am3) -- (lam3);
    \draw[->] (bm3) -- (lbm3);
    \draw[->] (bm3) -- (rbm3);

    %% bottom right tree
    %% \node[slave] (lr) at (left-root) {GP};
    \node[slave] (mr4) [below left  = 4.0cm and 0.0cm of right-root, label=right:GP] {6};
    %% [below left  = 0.5cm and 0.5cm of left-root, label=left:Q]
    \node[worker] (pm4) [below left  = 0.5cm and 0.5cm of mr4, label=left:pivot P]        {2};
    \node[slave] (qm4) [below right  = 0.5cm and 0.5cm of pm4, label=right:Q]  {4};
    \node[slave] (am4) [below left  = 0.5cm and 0.5cm of pm4, label=left:A]         {1};
    \node[worker] (bm4) [below left  = 0.5cm and 0.5cm of qm4, label=right:B]       {3};
    \node[slave] (cm4) [below right  = 0.5cm and  0.5cm of qm4, label=right:C]       {5};

    \node[] (lam4) [below left  = 0.5cm and  0.5cm of am4]        {};
    \node[] (lbm4) [below left  = 0.5cm and  0.5cm of bm4]        {};
    \node[] (rbm4) [below right  = 0.5cm and  0.5cm of bm4]        {};

    \draw[thick,->,gray,densely dotted] (mr4) -- node[black,pos=0.45,xshift=5pt] {\emph{D}} (qm4) ;
    \draw[thick,->] (mr4) to[out=-180,in=80] node [pos=0.3,below=1pt] {\emph{T}} (pm4);
    %%\draw[thick,->,gray,densely dotted] (qm) -- node [black,pos=0.2,below=-1pt] {\emph{S3}} (pm);
    %%\draw[thick,->] (qm) to[out=-105,in=105] (bm);
    \draw[->] (pm4) -- (am4);
    %%\draw[thick,->,gray,densely dotted] (pm) -- (bm);
    \draw[->] (pm4) -- (qm4);
    \draw[->] (qm4) -- (cm4);
    \draw[->] (qm4) -- (bm4);

    \draw[->] (am4) -- (lam4);
    \draw[->] (bm4) -- (lbm4);
    \draw[->] (bm4) -- (rbm4);

    %% right tree
    %%\node[slave] (rr) at (right-root) {GP};
    \node[slave] (rr5) [below left  = 0.0cm and 0.0cm of right-root, label=right:GP] {6};
    \node[worker] (pr5) [below left  = 0.5cm and 0.5cm of rr5, label=left:pivot P]        {2};
    \node[slave] (ar5) [below left  = 0.5cm and 0.5cm of pr5, label=left:A]        {1};
    \node[slave] (qr5) [below right  = 0.5cm and 0.5cm of pr5, label=right:Q]        {4};
    \node[worker] (br5) [below left  = 0.5cm and 0.5cm of qr5, label=left:B]        {3};
    \node[slave] (cr5) [below right  = 0.5cm and 0.5cm of qr5, label=right:C]        {5};

    \node[] (lar5) [below left  = 0.5cm and  0.5cm of ar5]        {};
    \node[] (lbr5) [below left  = 0.5cm and  0.5cm of br5]        {};
    \node[] (rbr5) [below right  = 0.5cm and  0.5cm of br5]        {};

    \draw[->] (rr5) -- (pr5);
    \draw[->] (pr5) -- (ar5);
    \draw[->] (pr5) -- (qr5);
    \draw[->] (qr5) -- (br5);
    \draw[->] (qr5) -- (cr5);

    \draw[->] (ar5) -- (lar5);
    \draw[->] (br5) -- (lbr5);
    \draw[->] (br5) -- (rbr5);

    \draw[ultra thick,<-] (ql1 -| pl1) -- (pl1) node[pos=0.3,below,sloped,rotate=90,xshift=-2mm] {+1};
    \draw[ultra thick,->] (pr5 -| qr5)  -- (qr5) node[pos=0.7,above,sloped,rotate=90,xshift=2mm] {-1};

    \draw[->,thick] ([xshift=1.2cm]ql1.east) -- ([xshift=-1.4cm]pr5.west) node[pos=0.5,above]{Right rotation};

    \node[] at (alignment-end -| qm2) {1. Q(L) Drops P and Takes B (S1)};
    \node[] at (alignment-end -| qm3) {2. P(R) Drops B and Takes Q (S2)};
    \node[] at (alignment-end -| mr4) {3. GP(L) Drops Q and Takes P (S3)};

  \end{scope}
\end{tikzpicture}
  \caption{Right rotation with red (white) and black (gray)
    nodes. Nodes/Roles (A, B, C, pivot P, Q, and GP) and keys (1, 2,
    3, \ldots, 6). P is promoted and Q is demoted. Nodes drop (D) and
    take (T) children.\label{fig.rotation}}
\end{figure}

\subsection{Tree Rotations\label{sec.rotations}} A common balancing
technique in binary trees are tree
rotations~\cite{Sleator:1986:RDT:12130.12143}. Tree rotations preserve
the order of nodes, but change the shape of the tree for
rebalancing. \figref{fig.rotation} shows an example for a right
rotation with Q as the root of the rotation, P as pivot (the left
child of Q), and GP as grandparent of P. The rotation increases the
height of the tree under the pivot by one and decrease the height of
the tree under the root by one. Additionally, it solves a red
violation between the pivot (\raisebox{-1.8pt}{\tikz{\node[draw,
      circle, minimum size=3mm, inner sep=0pt] {\smaller 2};}}) and B
(\raisebox{-1.8pt}{\tikz{\node[draw, circle, minimum size=3mm, inner
      sep=0pt] {\smaller 3};}}). A left rotation works vice versa. The
order of keys (1, 2, 3, $\ldots$, 6) and thus the order of nodes
remains unchanged.

For the right rotation, Q replaces its left child (the pivot) with B.
The pivot replaces its right child (B) with Q. The GP replaces its
left child (Q) with the pivot. The three nodes pass the ownership,
the parent relation, around. This leads to the tree rotating around the
GP.

\paragraph{Atomic Tree Rotations for \nvm{}}

For \nvm{}, the ownership changes are implemented with stores
(S1-S3). S1 for updating Q's left child, S2 for updating the pivot's
right child, and S3 for updating GP's left child. If the ownership
changes are performed partially, i.e, the pivot drops the link to B
and Q does not take ownership of B, we lose sub-trees and can create
cycles, as the following analysis shows:

\begin{tabular}{lllll}
  \emph{S1} and \emph{S3} & fail &$\rightarrow$& No link to pivot P. & \\
  \emph{S1} and \emph{S2} & fail &$\rightarrow$& No link to B. & Cycle: Q and pivot P. \\
  \emph{S2} and \emph{S2} & fail &$\rightarrow$& No link to Q.\\
  \emph{S1}       & fails     &$\rightarrow$& No link to pivot P.\\
  \emph{S3}       & fails     &$\rightarrow$& No link to Q.\\
  \emph{S2}       &  fails    &$\rightarrow$& No link to B. & Cycle: Q and pivot P.\\
\end{tabular}

GP, Q, and the pivot have to be updated atomically. Otherwise, the
tree will lose sub-trees and becomes invalid. Logging the insert
resp.\ delete request would not be sufficient. Shadowing would copy Q
and the pivot node, update them, and atomically update the pointer of
the grandparent pointing to the new pivot. In our approach, we copy
pointers to Q, the pivot, and B to the \redolog{} in a first step and
then update Q, the grandparent, and the pivot in a second transaction.

\section{Implementing binary trees with \mtx{}}\label{sec.rbrdma}

In the following, we describe how to implement \rbts{} (RBTs) in
\nvm{}. The major challenge is to convert the RBT insert and remove
algorithms into \sms{} with idempotent state transitions. For
\avlts{}, we use the same concepts. For didactic and brevity reasons,
we focus on RBTs in the following.

\subsection{Creating the \sm{}}

In the following, we outline how to create the \sm{}. For this paper,
we build them manually following the techniques described
below. Building a tool that automatically converts an algorithm into a
\sm{} with idempotent state transitions is beyond the scope of this paper.

As a first approximation, the imperative algorithm is converted into a
control-flow graph~\cite{Allen:1970:CFA:390013.808479} with basic
blocks. The nodes represent basic blocks resp.\ state transitions and
edges represent states. If a basic block is not idempotent, it has
overlapping read and write sets and has to be split into two. The
first block writes data into the \redolog{} and the second block
updates data in-place. This process has to be iterated until all state
transitions are idempotent.

For example, a tree rotation updates pointers in-place. Its
read and write sets overlap. The overwritten values, which can be
determined by alias analysis~\cite{aho1986compilers}, have to be
stored in the \redolog{} first before the actual rotation is performed.

While creating the \sm{} and its \redolog{}, we tried to find a
balance between convenience and the log's size. Minimizing the log's
size is equivalent to the register allocation
problem~\cite{pereira2006register}. An example for the flexibility in
design are double rotations. They could be expressed as two single
rotations (4 state transitions, space for 3 pointers in the log) or as
one double rotation (2 state transitions, space for 6 pointers in the
log). Trivial control-flow can often be hidden in state transitions,
like, e.g.,

\code{Direction dir = (it->key < key) ? Right : Left;}

\subsection{Insert\label{sec.fsminsert}}

\begin{figure}[t!]
  \centering
\begin{lstlisting}[numbers=left,stepnumber=1,numbersep=1ex,xleftmargin=3.5ex]
void insert(root, key, value) {
 if (root == nullptr) { // the first node
   root = new Node(key, value);              (*@ \label{fig.tdinsert.newnode2} @*)             // A1
 } else { // initialize pointers and iterators
   Node head;                                              // A2
   Node *it, *parent, *grand, *grandgrand = nullptr; (*@ \label{fig.tdinsert.helperbegin} @*)
   parent = &head;
   it = parent->right = root;
   Direction dir, last; dir = Left;                  (*@ \label{fig.tdinsert.helperend} @*)
   while(true) {
     if (it == nullptr) { // insert the new node here
       parent->dir = it = new Node(key, value);      (*@ \label{fig.tdinsert.newnode} @*)     // A3,A4
     } else if (isRed(it->left) and isRed(it->right)) {
       // recolor
       it->color = Red;  (*@ \label{fig.tdinsert.flipbegin} @*)                                 // A5
       it->left->color = it->right->color = Black; (*@ \label{fig.tdinsert.flipend} @*)
     }
     if (isRed(it) and isRed(parent) // need rotation?
       rebalance(grandgrand, grand, last); (*@ \label{fig.tdinsert.balance} @*)               // A6,A7,A8,A9,A10,A11
     if (it->key == key) break; // key exists already
     // traverse one level down the tree
     last = dir; dir = (it->key < key) ? Right : Left;     // A12
     grandgrand = grand; (*@ \label{fig.tdinsert.shufflebegin} @*)
     grand = parent; parent = it;
     it = it->dir; (*@ \label{fig.tdinsert.shuffleend} @*)
   }
   // update root
   root = head.right;  (*@ \label{fig.tdinsert.updateroot} @*)                                   // A13
 }
 // ensure the root is black
 root->color = Black;  (*@ \label{fig.tdinsert.maketherootblack} @*)                                   // A14
}
\end{lstlisting}
\caption{Top down insert based on~\citet{redblacktrees} and J. Walker.$^2$\label{fig.tdinsert}
The labels show the state of the \sm{} in \figref{fig.rbtree-fsm}.}
\end{figure}

\citet{redblacktrees} introduced top-down
approaches for inserting and removing key-value pairs for dichromatic
trees, see \figref{fig.tdinsert}. As they are top-down algorithms,
they do not require keeping a stack, but only keep a few pointers up
the tree---the iterators. They are mainly used for playing the
roles/anchors in tree rotations.  They also use a fake head node to simplify
corner-cases. We derived insert and remove (\figref{fig.tdinsert}
and~\ref{fig.tdremove}) from Julienne
Walker,$^2$
% same footnote as before.
%\footnote{http://www.eternallyconfuzzled.com/tuts/datastructures/jsw\_tut\_rbtree.aspx}
who uses the same concepts.  To avoid black violations, we insert the
new node as a red node at the bottom. However, inserting a red node
can yield red violations, which can be resolved by
promotions~\cite{TARJAN1983253}, i.e., color flip, single, and double
rotation.

Inserting a node into an empty tree from the clean state \emph{C} is
trivial (\emph{C} $\rightarrow$ \emph{A1} $\rightarrow$ \emph{C}).
Otherwise, we initialize helpers with the \textsf{init} transition
(lines~\ref{fig.tdinsert.helperbegin}--\ref{fig.tdinsert.helperend};
outgoing edges from \emph{A2}). If a leaf node is reached, we insert
the new node (line~\ref{fig.tdinsert.newnode}; \emph{A3} $\to$
\emph{A4}).  Otherwise, we might need to flip colors
(lines~\ref{fig.tdinsert.flipbegin}--\ref{fig.tdinsert.flipend}),
which can be done in one state transition, because setting a new color
is idempotent (outgoing edges of \emph{A5}).

The \texttt{rebalance} function (line~\ref{fig.tdinsert.balance} in
\figref{fig.tdinsert}) performs single or double rotations between the
grand and grand grandparent. Single rotations are converted into two
state transitions: one for logging and one for executing the rotation
(\figref{fig.rbtree-fsm}, \emph{A10} $\rightarrow$ \emph{A11}
$\rightarrow$ either \emph{A12} (to continue) or \emph{A13} (key was
found)). Double rotations need four state transitions accordingly
(\emph{A6} $\rightarrow$ \emph{A7} $\rightarrow$ \emph{A8}
$\rightarrow$ \emph{A9}).

Lines~\ref{fig.tdinsert.shufflebegin}--\ref{fig.tdinsert.shuffleend}
descend the iterators one level down the tree (outgoing edges
of \emph{A8}) to close the loop. In~\secref{sec.optzs}, we show how
to update the iterators with fewer state transitions and persist operations in some cases. Finally, the root is set and
colored black (lines
\ref{fig.tdinsert.updateroot}--\ref{fig.tdinsert.maketherootblack};
\emph{A13} $\to$ \emph{A14} $\to$ \emph{C}).

The log's size (4 cache lines of 64 bytes each) is independent of the
size of the tree (see \figref{fig.rbtree-fsm}). Its main components
are the key, the value, the iterator, the parent, and the
grandparent. In addition, it keeps some space for redo logging, e.g.,
\emph{Dir}, \emph{TmpNode}, and \emph{Sp}. The direction on the way
down the tree (\emph{Dir}). A temporary node for tree rotations
(\emph{TmpNode}). An anchor node on the way down the tree for remove
in RBT (\emph{Sp}). The majority of the space would also be needed for
non-persistent insert operation. Note that the log does not keep a
stack of size $\mathcal{O}(\log{n})$. Instead, it suffices to keep a
few pointers up the tree.

\subsection{Remove\label{sec.fsmremove}}

\begin{figure}[t!]
  \centering
\begin{lstlisting}[numbers=left,stepnumber=1,numbersep=1ex,xleftmargin=3.5ex]]
void remove(root, key) {
 if (root == nullptr) return; // empty tree
 // initialize pointers and iterators
 Node head;                                      (*@ \label{fig.tdremove.initvarsbegin} @*)        // C0
 Node *it, *parent, *grand, *found = nullptr;
 Direction dir = Right;
 it = &head;
 it->right = root;                               (*@ \label{fig.tdremove.initvarsend} @*)
 while (it->dir != nullptr) {
  // traverse one level down the tree
  Direction last = dir;                         (*@ \label{fig.tdremove.shufflebegin} @*)         // R1
  grand = parent;
  parent = it;
  it = it->dir;                                (*@ \label{fig.tdremove.shuffleend} @*)
  dir = (it->key < key) ? Right:Left;  // direction?
  found = (it->key == key) ? it:found; // found?
  if (not isRed(it) and not isRed(it->dir)) {
    if (isRed(it->(!dir))) { // single rotation
      parent = parent->last = single(it, dir);  (*@ \label{fig.tdremove.single} @*)         // R2,R3
    } else if (not isRed(it->(!dir))) {
      Node *s = parent->(!last);                          // R4
      if (s != nullptr) {
       Direction dir2 =
         (grand->right == parent)? Right : Left;
       if (not isRed(s->left) and not isRed(s->right)){
         // recolor
         parent->color = Black;  (*@ \label{fig.tdremove.flipbegin} @*)                        // R5
         s->color = Red;
         it->color = Red;                      (*@ \label{fig.tdremove.flipend} @*)
       } else if ((grand != nullptr) and
       not((grand==&head) and (dir2==Left))){
         // rotate?
         rebalance(grand, parent, s, last); (*@ \label{fig.tdremove.rotate} @*)             // R6,R7,R8,R9,R10,R11
         // recolor
         it->color = g->dir2->color = Red;  (*@ \label{fig.tdremove.colorbegin} @*)             // R12
         grand->dir2->left->color = Black;
         grand->dir2->right->color = Black; (*@ \label{fig.tdremove.colorend} @*)
 } } } } }
 if (found != nullptr) { // unlink and delete
   found->key = it->key; (*@ \label{fig.tdremove.deletebegin} @*)                                // R13,R14
   Direction dirL = (parent->right == it) ? Left:Right;
   Direction dirR = (it->left == nullptr) ? Right:Left;
   parent->dirL = it->dirR;
   delete it; (*@ \label{fig.tdremove.deleteend} @*)
 }
 // update root
 root = head.right;    (*@ \label{fig.tdremove.rootbegin} @*)                                  // R15
 root->color = Black; // ensure the root is black (*@ \label{fig.tdremove.rootend} @*)
}
\end{lstlisting}
\caption{Top down remove based on~\citet{redblacktrees} and J. Walker.$^2$\label{fig.tdremove} The labels show the state of the \sm{} in \figref{fig.avlt-fsm}.}
\end{figure}

While the algorithm for remove (\figref{fig.tdremove}) looks more
complex than for insert (\figref{fig.tdinsert}), its \sm{} in
\figref{fig.rbtree-fsm} is simpler than the one for inserting. The
control flow graph for remove is simpler and consists of a single
nested \texttt{if} statement. In each iteration, the \sm{} can jump
into exactly one place. In contrast, the inner part of the insert
algorithm contains two nested \texttt{if} statements. It allows jumping
into two cases resulting in a more complex \sm{}.

After the basic initialization,
lines~\ref{fig.tdremove.shufflebegin}--\ref{fig.tdremove.shuffleend}
move the iterators one level further down the tree. It uses the
even-odd optimization described in \secref{sec.optzs}, so that all
outgoing edges of \emph{R1} update the iterators. The single rotation
in line~\ref{fig.tdremove.single} uses \emph{R2} $\to$ \emph{R3}. The
color flip in
lines~\ref{fig.tdremove.flipbegin}--\ref{fig.tdremove.flipend} is
idempotent. Thus, a single state transition from \emph{R5}
suffices. The same applies to the color correction in
lines~\ref{fig.tdremove.colorbegin}--\ref{fig.tdremove.colorend}
represented by \emph{R12}.

\begin{figure*}[t!]
  \centering
  \begin{tikzpicture}[>=stealth]
  \coordinate (origin) at (0,0);
  \scriptsize

  %%\sffamily%%  \smaller\smaller
  \let\sizebox\relax
  \newlength{\sizebox}\setlength{\sizebox}{3.0mm}

  \tikzstyle{box}=[draw,inner sep=0pt,text width=\sizebox,minimum size=\sizebox,minimum height=4mm,text height=1.5ex,text depth=.25ex]

  \begin{scope}[start chain=p going right]
    \node [node distance= 0mm,on chain=p] (alignment) {};
    \node [node distance= 0mm,on chain=p] (tree-chain) {};
  \end{scope}

  %% protocol
  \begin{scope}[
      start chain=alignment going below,
      start chain=tree-chain going below,
      node distance=2mm]

    %% start time lines
    \foreach \x in {alignment, tree-chain}
    \chainin (\x) [on chain=\x];
  \end{scope}

    %% legend
    \node[node distance =   58mm, on chain=alignment] (alignment-legend1) {};
    \node[node distance =   5mm, on chain=alignment] (alignment-legend2) {};
    \node[node distance =   4mm, on chain=alignment] (alignment-legend) {};

    \input{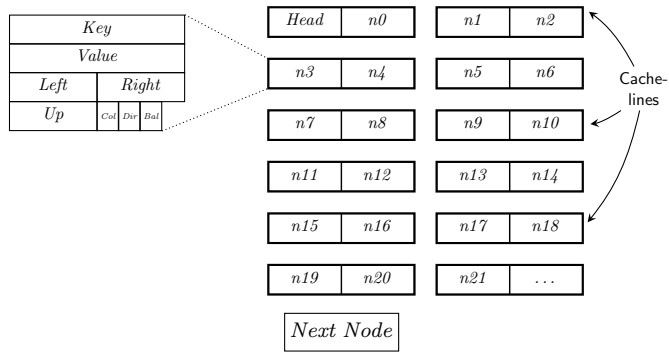}

    \begin{scope}[local bounding box=treearray]
      \pic[anchor=north west]  at (tree-chain) {tree2={.8}};
    \end{scope}

    %% \node[] at (alignment-legend -| treearray.south) {Tree};

    \tikzstyle{boxbox}=[draw,inner sep=0pt,text width=1.5cm,minimum width=1.5cm,minimum height=5mm,text height=1.5ex,text depth=.25ex]

   \node[boxbox,align=center] at (alignment-legend2 -| treearray.south) {\larger\emph{Next Node}};

\end{tikzpicture}
  \caption{Tree-nodes occupy 31~bytes aligned to 32~bytes,
    tree-pointers need 4~bytes, keys and values require 8~bytes each,
    and directions (\emph{Dir}), the balance (\emph{Bal}), and colors
    (\emph{Col}) take one byte each. The \textsf{Next Node} fills
    7~bytes aligned to 8~bytes. \label{fig.common-tree}}
\end{figure*}

The balance operation in line~\ref{fig.tdremove.rotate} either
executes a single rotation (\emph{R6} $\to$ \emph{R7}) or a
double rotation (\emph{R8} $\to$ \emph{R9} $\to$
\emph{R10} $\to$ \emph{R11}).

\subsection{Common state for Trees on \nvm{}}\label{sec.archtree}

The tree (\figref{fig.common-tree}) is represented as a fixed-length
array of tree nodes. Pointers to nodes are thus indices into the
array, so that the pointers remain valid when the array is mapped to a
different base address after a crash. The first element is reserved
for the special head node in the insert and remove algorithms. The
node data-structure is shared by \trees{}. The \texttt{Col} (the
color) field is used for RBTs. \texttt{Up} (up pointer \js{pointer to
  the parent?}), \texttt{Dir} (direction from parent), and
\texttt{Bal} (the balance) are exclusively used for
\avlts{}. \texttt{Up} and \texttt{Dir} are needed to walk the tree
back up again, see \secref{sec.avlrees}.

The programming model for \nvm{} requires the tree to be a
fixed-length array~\cite{sniapm}. The size of mmapped files for \nvm{}
cannot change. Increasing the tree's size further would require
adding additional fixed-length arrays. The index type would need to be
updated accordingly (from IdxType to (ArrayIdxType, IdxType)).

To support allocating new tree nodes after a crash, we use the
\emph{Next Node} structure---a \texttt{uint56\_t}. It stores the
index of the next free array element. Note though, we did not
implement garbage collection. Despite a client releasing a node, it
cannot be used anymore. We only allocate from the head and do not
maintain free lists.

\nvm{}-aware garbage collection and dynamic node allocation is
provided with Makalu~\cite{Bhandari:2016:MFR:2983990.2984019},
nvm\_malloc~\cite{conf/vldb/SchwalbBFDP15}, and
PAllocator~\cite{palloc}. They could be used for our tree data
structure. The state and next node variable have to be updated with
atomic CAS operations. All other updates can use non-atomic stores,
because we can do redo logging.

\subsection{State for \rbts{} on \nvm{}}\label{sec.archrbt}

\begin{figure*}[t!]
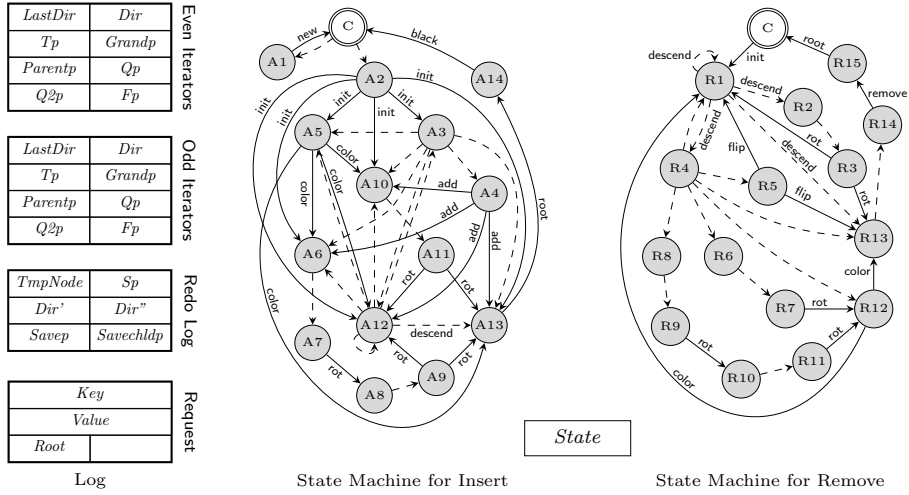

  \centering
  \begin{tikzpicture}[>=stealth]
  \coordinate (origin) at (0,0);
  \scriptsize

  %%\sffamily%%  \smaller\smaller
  \let\sizebox\relax
  \newlength{\sizebox}\setlength{\sizebox}{3.0mm}

  \tikzstyle{box}=[draw,inner sep=0pt,text width=\sizebox,minimum size=\sizebox,minimum height=4mm,text height=1.5ex,text depth=.25ex]

  \begin{scope}[start chain=p going right]
    \node [node distance= 0mm,on chain=p] (alignment) {};
    \node [node distance= 0mm,on chain=p] (tree-chain) {};
    \node [node distance= 0mm,on chain=p] (logl-chain) {};
    \node [node distance=11mm,on chain=p] (logm-chain) {};
    \node [node distance=35mm,on chain=p] (sm-insert-chain) {};
    \node [node distance=50mm,on chain=p] (sm-remove-chain) {};
  \end{scope}

  %% protocol
  \begin{scope}[
      start chain=alignment going below,
      start chain=tree-chain going below,
      start chain=logl-chain going below,
      start chain=logm-chain going below,
      start chain=logr-chain going below,
      start chain=sm-insert-chain going below,
      start chain=sm-remove-chain going below,
      node distance=2mm]

    %% start time lines
    \foreach \x in {alignment, tree-chain, sm-insert-chain, sm-remove-chain,
      logl-chain, logm-chain, logr-chain}
    \chainin (\x) [on chain=\x];
  \end{scope}

    %% legend
    \node[node distance =   58mm, on chain=alignment] (alignment-legend1) {};
    \node[node distance =   0mm, on chain=alignment] (alignment-legend2) {};
    \node[node distance =   4mm, on chain=alignment] (alignment-legend) {};

    %% log iterators
    \node[node distance =    0mm, on chain=logl-chain] (iterators-left) {};
    \node[node distance =    17mm, on chain=logl-chain] (iterators-right) {};

    %% log redo
    \node[node distance =  17mm, on chain=logl-chain] (redo-log) {}; %% FIXME

    %% log log
    \node[node distance =  14mm, on chain=logl-chain] (log-log) {};

    \input{common.tex}

    %%\begin{scope}[local bounding box=treearray]
    %%  \pic[anchor=north west]  at (tree-chain) {tree2={.8}};
    %%\end{scope}

    \begin{scope}[local bounding box=its-left]
      \pic at (iterators-left) {iterators={0.8}};
    \end{scope}

    \begin{scope}[local bounding box=its-right]
      \pic at (iterators-right) {iterators={0.8}};
    \end{scope}

    \begin{scope}[local bounding box=redo-bb]
      \pic at (redo-log) {redo={0.8}};
    \end{scope}

    \begin{scope}[local bounding box=log-bb]
      \pic at (log-log) {log={0.8}};
    \end{scope}

    \begin{scope}[local bounding box=bb-insert]
      \normalsize\node[anchor=north,inner sep=0pt] (russell) at (sm-insert-chain)
           {\input{fsm-insert2}};
    \end{scope}

    \begin{scope}[local bounding box=bb-remove]
      \normalsize\node[anchor=north,inner sep=0pt] (russell) at (sm-remove-chain)
           {\input{fsm-remove3}};
    \end{scope}

    \begin{scope}
      \begin{pgfonlayer}{background}
        \node[fit=(its-left)] (cl-its-left) {};
        \node[fit=(its-right)] (cl-its-right) {};
        \node[fit=(redo-bb)] (cl-redo) {};
        \node[fit=(log-bb)] (cl-log) {};
      \end{pgfonlayer}
    \end{scope}

    %% labels: cache
    \node[rotate=270,yshift=1.5mm] at (cl-its-left.east) {\textsf{Even
    Iterators}};
    \node[rotate=270,yshift=1.5mm] at (cl-its-right.east) {\textsf{Odd
    Iterators}};

    \node[rotate=270,yshift=1.5mm] at (cl-redo.east) {\textsf{Redo Log}};
    %% \node[anchor=north] at (cl-log.south) {Log};

    \node[rotate=-90,yshift=1.5mm] at (cl-log.east) {\textsf{Request}};
    %% \node[rotate=270,yshift=1.5mm] at (cl-its-right.east) {\smaller Iterators};

    %% labels
    \node[] at (alignment-legend -| bb-remove.south) {State Machine for Remove};
    \node[] at (alignment-legend -| bb-insert.south) {State Machine for Insert};

    %% \node[] at (alignment-legend -| treearray.south) {Tree};
    \node[] at (alignment-legend -| log-bb.south) {Log};

    \tikzstyle{boxbox}=[draw,inner sep=0pt,text width=1.5cm,minimum width=1.5cm,minimum height=5mm,text height=1.5ex,text depth=.25ex]

    %% \node[boxbox,align=center] at (alignment-legend2 -| treearray.south) {\larger\emph{Next Node}};

    \node[boxbox,align=center,xshift=3mm] at (alignment-legend2 -| bb-insert.south east) {\larger\emph{State}};

\end{tikzpicture}
  \caption{\nvm{} data structures and \sms{} for insert and remove for
    \rbts{} with 15 (insert) and 16 (remove) states
    (\textsf{rot}=rotation, \textsf{init}=initialize variables,
    \textsf{add}=insert new node, \textsf{remove}=remove a node,
    \textsf{descend}=one step down the tree, \textsf{root}=update
    root, \textsf{color}=recolor nodes, \textsf{black}=color the root
    black). Dashed edges flush to the log. The \textsf{State} fills
    7~bytes aligned to 8~bytes.\label{fig.rbtree-fsm}}
\end{figure*}

The main data structures to manage an RBT are stored in \nvm{}: the
tree, the log, the state variable, and the next node
(\figref{fig.common-tree} and \figref{fig.rbtree-fsm}). For
concurrency control, we additionally maintain a multi-reader
single-writer lock in volatile memory (see \secref{sec.locks}).

\ts{Providing parallel access to the state machine does not work at
  all. Each thread would have to have its own state machine instance.}

\ts{For low-radix trees, concurrent writers would not work on disjoint
  sets. Insert and remove always touch large parts of the tree. For B+
  trees, insert and remove work on leaves and thus multi-threading
  shows higher gains. }

The log is structured into four parts, each fitting into a single
cache line: data for the root node and the request to perform, data to
redo the current state transition, and even and odd data structures
(see \secref{sec.optzs} for more details on this) for traversing the
tree.

\subsection{State for \avlts{} on \nvm{}}\label{sec.archavlt}

\begin{figure*}[t!]
  \centering
  \begin{tikzpicture}[>=stealth]
  %%\sffamily%%
  \scriptsize
  \let\sizebox\relax
  \newlength{\sizebox}\setlength{\sizebox}{3.0mm}

  \tikzstyle{logbox1}=[draw,inner sep=0pt,text width=18\sizebox,minimum width=20\sizebox+3*\pgflinewidth,minimum height=6mm,text height=1.5ex,text depth=.25ex]

  \tikzstyle{logbox2}=[draw,inner sep=0pt,text width=9\sizebox,minimum width=10\sizebox+2*\pgflinewidth,minimum height=6mm,text height=1.5ex,text depth=.25ex]

  \tikzstyle{logbox8}=[draw,inner sep=0pt,text width=2.15\sizebox,minimum width=2.5\sizebox+\pgflinewidth,minimum height=6mm,text height=1.5ex,text depth=.25ex]

    %%\begin{scope}[start chain=lines going {below=of \tikzchainprevious.south west}, node distance=4mm]
    \begin{scope}[start chain=lines going right]
      \node[node distance= 0mm, on chain=lines, anchor=west] (alignment-anchor) {};
      \node[node distance= 0mm, on chain=lines, anchor=west] (log-anchor) {};
      \node[node distance=38mm, on chain=lines, anchor=west] (sm1-anchor) {};
      \node[node distance=42mm, on chain=lines, anchor=west] (sm2-anchor) {};
    \end{scope}

    \begin{scope}[align=center, start chain=logchain going below]
      \chainin (log-anchor) [on chain=logchain];
      \node[on chain=logchain, anchor=west] (cl1) {};
      \node[node distance=30mm, on chain=logchain, anchor=west] (cl2) {};
    \end{scope}

    \begin{scope}[align=center, start chain=alignmentchain going below]
      \chainin (alignment-anchor) [on chain=alignmentchain];
      \node[node distance=70mm, on chain=alignmentchain, anchor=west] (alignment) {};
    \end{scope}

    \begin{scope}[align=center, node distance=-\pgflinewidth,every node/.style={scale=0.6}]
      \node[logbox1] (r14)  at (cl1)   {\larger\larger\emph{Key}};
      \node[logbox1] (r15) [below= of r14] {\larger\larger\emph{Value}};

      \node[logbox2] (r16) [below= of r15.south west, anchor=north west] {\larger\larger\emph{Root}};
      \node[logbox2] (r17) [right= of r16] {\larger\larger\emph{Root2}};

      \node[logbox2] (r18) [below= of r16.south west,anchor=north west] {\larger\larger\emph{P}};
      \node[logbox2] (r19) [right= of r18] {\larger\larger\emph{Q}};

      \node[logbox2] (r20) [below= of r18.south west,anchor=north west] {\larger\larger\emph{S}};
      \node[logbox2] (r21) [right= of r20] {\larger\larger\emph{T}};

      \node[logbox2] (r22) [below= of r20.south west,anchor=north west] {\larger\larger\emph{NextP}};
      \node[logbox2] (r23) [right= of r22] {\larger\larger\emph{NextS}};

      \node[logbox2] (r24) [below= of r22.south west,anchor=north west] {\larger\larger\emph{NextT}};
      \node[logbox2] (r25) [right= of r24] {\larger\larger\emph{SP}};

      \node[logbox2] (r26) [below= of r24.south west,anchor=north west] {\larger\larger\emph{N}};
      \node[logbox2] (r27) [right= of r26] {\larger\larger\emph{NN}};

      \draw[thick] (r14.north west) rectangle (r27.south east);

       {\larger\larger \node[node distance=4mm] (cacheline) [left= of r26] {\textsf{Cache-}\\\textsf{lines}};}

    \end{scope}

    \begin{scope}[align=center, node distance=-\pgflinewidth,every node/.style={scale=0.6}]
      \node[logbox2] (r28) at ([xshift=-10mm]cl2) {\larger\larger\emph{Save}};
      \node[logbox2] (r29) [right= of r28] {\larger\larger\emph{SaveChild}};

      \node[logbox8] (r30) [below= of r28.south west, anchor=north west] {\larger\larger\emph{Inc}};
      \node[logbox8] (r31) [right= of r30] {\larger\larger\emph{Bal}};
      \node[logbox8] (r32) [right= of r31] {\larger\larger\emph{Bal2}};
      \node[logbox8] (r33) [right= of r32] {\larger\larger\emph{Bal3}};
      \node[logbox2] (r33b) [right= of r33] {\larger\larger\emph{It}};

      \node[logbox2] (r34) [below= of r30.south west, anchor=north west] {\larger\larger\emph{OldIt}};
      \node[logbox2] (r35) [right= of r34] {\larger\larger\emph{NewItDown}};

      \node[logbox2] (r36) [below= of r34.south west, anchor=north west] {\larger\larger\emph{NewIt}};
      \node[logbox2] (r37) [right= of r36] {\larger\larger\emph{NewItUp}};

      \node[logbox2] (r38) [below= of r36.south west, anchor=north west] {\larger\larger\emph{OldNewIt}};
      \node[logbox2] (r39) [right= of r38] {\larger\larger\emph{OldNewItUp}};

      \node[logbox2] (r40) [below= of r38.south west, anchor=north west] {\larger\larger\emph{SaveParent}};
      \node[logbox2] (r41) [right= of r40] {\larger\larger\emph{Heir}};

      \node[logbox2] (r42) [below= of r40.south west, anchor=north west] {\larger\larger\emph{OldHeir}};
      \node[logbox2] (r43) [right= of r42] {\larger\larger\emph{Foo}};

      \node[logbox8] (r44) [below= of r42.south west, anchor=north west] {\larger\larger\emph{Foo}};
      \node[logbox8] (r45) [right= of r44] {\larger\emph{Done}};
      \node[logbox8] (r46) [right= of r45] {\larger\larger\emph{Top}};
      \node[logbox8] (r47) [right= of r46] {\smaller\emph{OldTop}};

      \node[logbox8] (r48) [right= of r47] {\larger\larger\emph{Dir}};
      \node[logbox8] (r49) [right= of r48] {\larger\larger\emph{Dir2}};
      \node[logbox8] (r50) [right= of r49] {\larger\larger\emph{Dir3}};
      \node[logbox8] (r51) [right= of r50] {\emph{FooDir}};

      \draw[thick] (r28.north west) rectangle (r51.south east);
    \end{scope}

    \begin{scope}[local bounding box=bb-insert]
      \normalsize\node[anchor=north,inner sep=0pt] at ([yshift=-10mm]sm1-anchor)
           {\input{avl-insert-sm}};
    \end{scope}

    \begin{scope}[local bounding box=bb-remove]
      \normalsize\node[anchor=north,inner sep=0pt] at ([yshift=-10mm]sm2-anchor)
           {\input{avl-remove-sm}};
    \end{scope}

    \draw[->,shorten >=1mm] (cacheline) to[bend left=13] (r20.north west);
    \draw[->,shorten >=1mm] (cacheline) to[bend right=13] (r36.west);

    \node[] at (alignment -| cl1) {Log};
    \node[] at (alignment -| sm1-anchor) {State Machine for Insert};
    \node[] at (alignment -| sm2-anchor) {State Machine for Remove};

    \tikzstyle{boxbox}=[draw,inner sep=0pt,text width=1.5cm,minimum width=1.5cm,minimum height=5mm,text height=1.5ex,text depth=.25ex]

    \node[boxbox,align=center] at ([yshift=7mm]alignment -| bb-insert.south east) {\larger\emph{State}};

\end{tikzpicture}
  \caption{\nvm{} data structures and state machines for insert and
    remove for \avlts{} with 24 and 35 states. Dashed edges flush to
    the log. The \textsf{State} fills 7~bytes aligned to
    8~bytes.\label{fig.avlt-fsm} \ts{in insert A3, A5, and A6 are
      missing, they are needed for wo look-ahead}}
\end{figure*}

The state for \avlts{} is similar to \rbts{}. While the log for \rbts{}
occupies four cache lines, the log for \avlts{} fits into two cache
lines. The overhead can mostly be accounted for the optimizations,
see \secref{sec.optzs}, and the cache lines are only partially
filled while the two cache lines for \avlts{} are completely
filled. Spreading the data over more cache lines might improve
performance further by reducing correlated cache misses, but such
optimizations are beyond the scope of this paper.

\section{Remote Persisting with a Software Agent\label{sec.sa}}

Clients efficiently access our \trees{} with passive target
communication using RDMA. All buffers used for such communication have
to be registered with the network card, which creates \emph{remote
  keys}. Clients have to use remote keys to access the registered address
range with RDMA operations. Revoking the key would deny the client
to access the address space. To gain access to remote keys for address
ranges with \infiniband{}, clients have to establish a connection and
query a \sa{} that we run on the server node. For each connection
request by a client, the \sa{} creates two client-specific memory
registrations: (a) $\mathit{mr}_d$ and (b) $\mathit{mr}_{\mathit{nv}}$
with read and write privileges.

Memory registrations allow revoking the access privileges of
individual users. If a client starts a failure detector on $c$ and the
failure detector claims $c$ to be failed, the client asks the \sa{} to
revoke access privileges of $c$, i.e., it invalidates its remote
key. This is a way to ensure client $c$ cannot access the data any
longer.  The client can then safely release locks held by $c$. After a
power loss, the DRAM region will be initialized with zeros. The locks
lose their value as the remote keys became invalid and can safely be
reset to not taken. \citet{poke2015dare} also use memory registration
and \infiniband{}'s QP mechanism to handle access rights, but fail to
realize that locks, see \secref{sec.locks}, or atomic operations are
required to prevent data-races.

As there is no hardware support for remote data persisting yet, the
\sa{} also provides cache flushing as a service.

\section{Optimizations\label{sec.optzs}}

\figref{fig.flush2} shows the epochs by category for inserting
resp.\ removing $10^{7}$ keys into \rbts{}. There are $\mathcal{O}(1)$ single and
double rotations per operation on average and one color flip as can be
expected~\cite{TARJAN1983253}. Some categories reflect necessary steps
for starting resp.\ completing operations, e.g., removing the found
node, updating the root, initializing the variables for the search,
and flushing the current command. While the cost for all operations is
in $\mathcal{O}(\log{n})$, the number of state transitions are
dominated by shuffling the iterators for tree traversal.

As we cannot update the iterators in place, we need to use the
\redolog{}. The canonical approach requires for each loop iteration
two state transitions (4 epochs). In the first, it flushes the old
iterators to the log. In the second, it updates the new iterators to go
one level deeper into the tree. Instead, we implemented an even-odd
scheme. All iterators are stored twice to have disjoint read and write
sets. In even rounds through the loop, the first set is written. In
odd rounds, the second set is written. Thus, we need only one state
transition and the previous iterators form the \redolog{} to go one
level down the tree. A set of iterators fits into a single cache line,
which supports the even-odd scheme and reduces flush costs. We use one
bit of the state variable to store whether we are in an even or odd
round:

\noindent
\begin{lstlisting}
struct SV {bool:1 Even{1},uint56_t:55 State{0}} EvenState;
\end{lstlisting}

\begin{figure}[h!]
  \centering
  \includegraphics{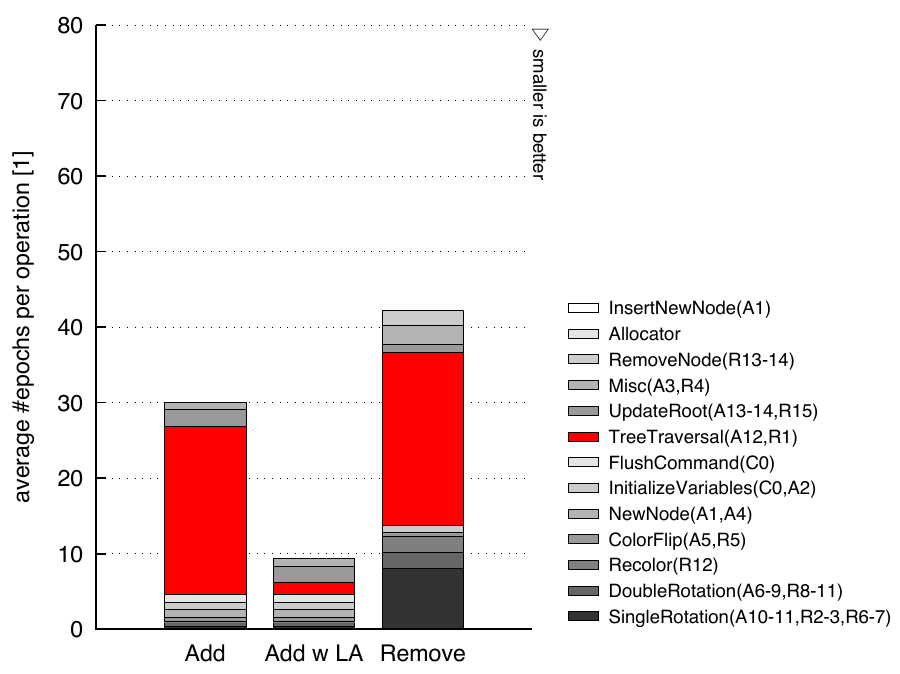}
  \caption{Average number of epochs for inserting resp.\ removing
    $10^7$ keys in random order in \rbts{}.\label{fig.flush2}}
\end{figure}

\begin{figure}[h!]
  \centering
  \includegraphics{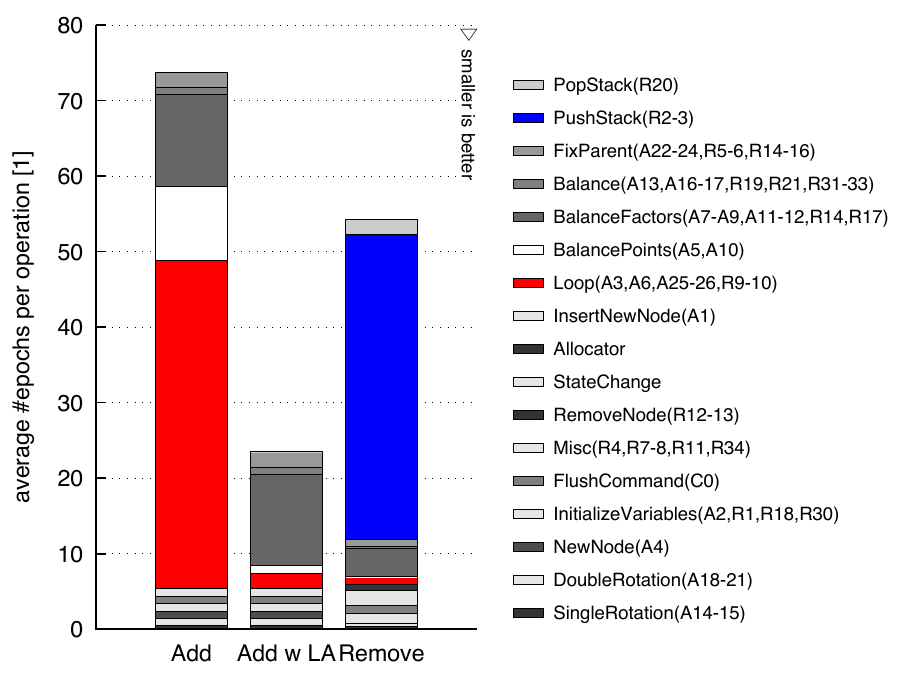}
  \caption{Average number of epochs for inserting resp.\ removing
    $10^7$ keys in random order in \avlts{}.\label{fig.flush2-avlt}}
\end{figure}

If the next loop iteration does not perform any rotations or
recoloring, it can be skipped. It does not change the tree. For
inserts, we added \emph{unlimited look-ahead} (LA): instead of going
exactly one level down into the tree in each iteration, we progress
the iterators directly to the next level that needs balancing or
recoloring.

\figref{fig.flush2} shows the average number of epochs, i.e., flushes
for inserting resp.\ removing $10^{7}$ keys in random
order~\cite{Matsumoto:1998:MTE:272991.272995} into RBTs. Insert and
remove are dominated by the tree traversal, but unlimited look-ahead
(LA) almost eliminates the costs. \figref{fig.flush2-avlt} shows the
transitions for \avlts{}. Add is dominated by the loop walking down
the tree. Look-ahead does not iterate through the loop. Remove is
dominated by push stack, which stores in nodes the pointers to their
parents. Again \avlts{} are more expansive than RBTs. In both
experiments, keys are inserted on average in depth 21. The final
\rbt{} has a depth of 29 and the \avlt{} has a depth of 28. It is
balanced more strictly.

The \avlt{} insert algorithm~\cite{Knuth:1998:ACP:280635} is split
into two phases. First, it walks down the tree and searches the parent
of the new node. On the way down, it saves two balancing points. In
the second phase, it inserts the new node and uses the balancing
points to rebalance the tree. The first loop can be replaced by two
state transitions, because it is almost side-effect free---except for
storing the balancing points. The approach is similar to the
look-ahead for RBTs.

The \avlt{} remove algorithm is more challenging, because the loops
walking down the tree store parent pointers in the nodes. Skipping
loop iterations is challenging. The last loop walks up the tree and
balances it. There are fewer opportunities for skipping iterations.

\section{Implementation Details\label{sec.impldetails}}

The programming model for \nvm{}~\cite{sniapm} maps files into memory
using \texttt{mmap}, which provides direct access (DAX) to the
NVDIMM. Mapping a file again after a power failure may yield a
different base address. So, all memory accesses have to be explicitly
adjusted to the corresponding address range. This is necessary to
consistently access the same data after a power
loss. Makalu~\cite{Bhandari:2016:MFR:2983990.2984019} is a persistent
heap manager, which hides this problem from users.

\medskip

For simplicity, we placed the log structure and the tree, an array of
nodes, into different files. There are pointers between nodes and
between the log and the nodes, which have to be adjusted to the base
addresses. The following assignment is completely handled by the
compiler.

\begin{lstlisting}
log->root->left->color = log->save->right->color;
\end{lstlisting}

There are no means to either control or adapt the intermediate memory
accesses. For \nvm{}, we need to control all memory accesses. We have
to map and adjust them to their respective locations in mmapped address
ranges. For remote access, we need to know all addresses and intermediate
steps to invoke put and get calls accordingly.

\begin{center}
\begin{tabular}{l|l}
expression & mmapped file \\ \hline
\lstinline|log->root| & log file\\
\lstinline|log->root->left| & tree file\\
\lstinline|log->root->left->color| & tree file\\
\lstinline|log->save| & log file\\
\lstinline|log->save->right| & tree file\\
\lstinline|log->save->right->color| & tree file\\
\end{tabular}
\end{center}

\noindent
The only feasible solution is to manage memory accesses by the user
instead of the compiler. Thus, for memory accesses and persist
operations, we wrote our own embedded domain specific language (DSL)
based on expression templates~\cite{citeulike:2443933}.

\medskip

The DSL provides a declarative language for describing memory
addresses including all intermediate steps---the path. Instead of
using expressions such as \lstinline|log->root->left->color|, we assign types to each memory
address, e.g.,\\
\lstinline|ColorInNode<LeftInNode<RootInLog>> address = {log};|.
Each memory access can be seen as a path of intermediate
memory accesses. The DSL allows programmers to describe memory
accesses and persist operations declaratively while ignoring the
peculiarities of the underlying programming model. For each memory
access, the runtime maps the request to the corresponding object (log,
nodes, next node, and state variable) and its associated mapped
file. The access is applied to the address space with the offset
adjusted accordingly.  This allows development on machines with and
without \nvm{} and facilitates remote access. For local development,
we simply use \texttt{malloc} to simulate mmapped files. For machines
with \nvm{}, we rely on the PMDK for mmapping. For remote access, we
use PMDK to mmap files and UCX~\footnote{http://www.openucx.org} for
communication with RDMA. Additionally, our DSL allows us to
transparently experiment with different cache flushing and caching
strategies. The runtime can map persist operations to the different
persist operations described in \secref{sec.howtopersist}. For remote
access, we can cache the results of gets.

\medskip

The following updates the key and value in the log:
\begin{lstlisting}
WriteOp<typename LogAddress::Key, KeyValue> W1 =
        {LogAddress::Key(log), KeyValue(key)};
WriteOp<typename LogAddress::Value, ValueValue> W2 =
        {LogAddress::Value(log), ValueValue(value)};
flushOp(W1, W2);
\end{lstlisting}

\noindent
\lstinline|WriteOp| and \lstinline|flushOp| are the customization
points. Each \lstinline|WriteOp| has a source and a destination memory
location. It performs a read and a write. \lstinline|flushOp| executes
all write operations. As memory locations for read and write are
described by paths, they have to be evaluated first. They may require
a sequence of plain loads for local access or gets for remote access. Each
\lstinline|WriteOp| could execute the assignment followed by a
\texttt{clwb} and the \lstinline|flushOp| invokes an \texttt{sfence}
to finish the epoch. For our tree code, the largest epoch (in AVLT
remove) invokes \lstinline|WriteOp| nine times.

\section{Multi-Reader Single-Writer Locks \\with
  $f$-fairness over RDMA\label{sec.locks}}

Data structures with concurrent writers can show greatly reduced
performance~\cite{scherer2005advanced}. Thus, we use multi-reader
single-writer (mrsw) locks~\cite{Courtois:1971:CCL:362759.362813} to
limit the number of concurrent writers. It is sufficient to store the
mrsw-lock in DRAM, because after power losses remote keys become
invalid (cmp.~\secref{sec.sa}).

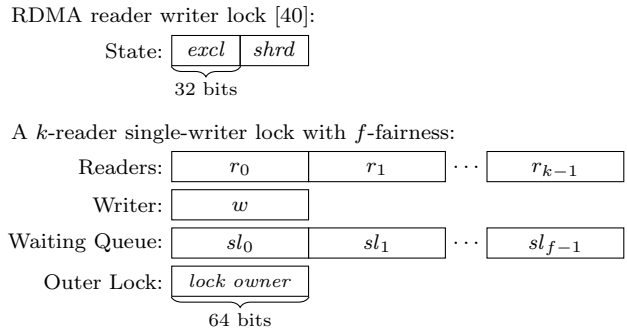
\begin{figure}[h!]
\centering\begin{tikzpicture}[>=stealth]
  %%\sffamily%%
  \smaller
  \let\sizebox\relax
  \newlength{\sizebox}\setlength{\sizebox}{3.0mm}

  \tikzstyle{box}=[draw,inner sep=0pt,text width=\sizebox,minimum size=\sizebox,minimum height=4mm,text height=1.5ex,text depth=.25ex]
  \tikzstyle{box2}=[draw,inner sep=0pt,text width=2\sizebox,minimum width=3\sizebox,minimum height=4mm,text height=1.5ex,text depth=.25ex]
  \tikzstyle{box4}=[draw,inner sep=1pt,text width=5\sizebox,minimum size=6\sizebox,minimum height=4mm,text height=1.5ex,text depth=.25ex]

  \begin{scope}[start chain=lines going {below=of \tikzchainprevious.south west}, node distance=4mm]
    \node[on chain=lines,anchor=west] {RDMA reader writer lock~\cite{narravula2007high}:};
    \node[node distance=2mm,on chain=lines,anchor=west] (simple-anchor) {};
    \node[node distance=10mm,on chain=lines,anchor=west] {A $k$-reader single-writer lock with $f$-fairness:};
    \node[node distance=2mm,on chain=lines,anchor=west] (ours-ranchor) {};
    \node[on chain=lines,anchor=west] (ours-wanchor) {};
    \node[on chain=lines,anchor=west] (ours-wqanchor) {};
    \node[on chain=lines,anchor=west] (ours-olanchor) {};
  \end{scope}

  \begin{scope}[align=center, node distance=-\pgflinewidth, start chain=simple going right]
    \chainin (simple-anchor) [on chain=simple];
    \node[box2,node distance=20mm,on chain=simple] (r0) {\emph{excl}};
    \node[box2,on chain=simple] (r1) {\emph{shrd}};
    \node[anchor=east] at (r0.west) {State:};
  \end{scope}

  %% readers
  \begin{scope}[align=center, node distance=-\pgflinewidth, start chain=oursreader going right]
    \chainin (ours-ranchor) [on chain=oursreader];
    \node[box4,node distance=20mm,on chain=oursreader] (or0) {$r_0$};
    \node[box4,on chain=oursreader] (or1) {$r_1$};
    \node[on chain=oursreader]  {$\ldots$};
    \node[box4,on chain=oursreader]  {$r_{k-1}$};
    \node[anchor=east] at (or0.west) {Readers:};
  \end{scope}

  %% writer
  \begin{scope}[align=center, node distance=-\pgflinewidth, start chain=ourswriter going right]
    \chainin (ours-wanchor) [on chain=ourswriter];
    \node[box4,node distance=20mm,on chain=ourswriter] (ow) {$w$};
    \node[anchor=east] at (ow.west) {Writer:};
  \end{scope}

  %% waiting queue
  \begin{scope}[align=center, node distance=-\pgflinewidth, start chain=ourswq going right]
    \chainin (ours-wqanchor) [on chain=ourswq];
    \node[box4,node distance=20mm,on chain=ourswq] (or0) {$\mathit{sl}_0$};
    \node[box4,on chain=ourswq] (or1) {$\mathit{sl}_1$};
    \node[on chain=ourswq]  {$\ldots$};
    \node[box4,on chain=ourswq]  {$\mathit{sl}_{f-1}$};
    \node[anchor=east] at (or0.west) {Waiting Queue:};
  \end{scope}

  %% outer lock
  \begin{scope}[align=center, node distance=-\pgflinewidth, start chain=oursol going right]
    \chainin (ours-olanchor) [on chain=oursol];
    \node[box4,node distance=20mm,on chain=oursol] (ol) {\smaller\emph{lock owner}};
    \node[anchor=east] at (ol.west) {Outer Lock:};
  \end{scope}

  \draw [decorate,decoration={brace,amplitude=4pt}] (r0.south east) -- (r0.south west) node [black,midway,below=3pt] {\smaller 32 bits};

  \draw [decorate,decoration={brace,amplitude=4pt}]
  (ol.south east) -- (ol.south west) node [midway,below=3pt]
        {\smaller 64 bits};

\end{tikzpicture}

%\bibliographystyle{ACM-Reference-Format}
%\bibliography{lock-types.bib}
%
\caption{State for reader-writer locks with RDMA.\label{fig.lockdesigns}}
\end{figure}

Two distributed lock servers based on RDMA were developed
by~\citet{narravula2007high} and \citet{DBLP:journals/corr/ChungZ15}
with similar designs.  They both use atomic CAS and fetch\&add to
update the lock data structure (see top of
\figref{fig.lockdesigns}). In exclusive mode, the \emph{excl} field
holds the client's process id (pid)~\cite{Armstrong:2013:PES:2566708}
holding the lock.  In shared mode, the \emph{shrd} field holds the
number of clients sharing the lock. The holders of the shared lock are
anonymous, which hinders to identify failed lock holders and to get
the lock back into exclusive mode. Thus, these designs are not well
prepared for client failures. They also cannot guarantee fairness,
because readers can always starve any writer trying to acquire the
write-lock. \citet{fompi-mpi3-one-sided} designed a similar lock for
one-sided communication in MPI, but here even the holder of the write
lock is anonymous. However, one could argue that as of today the
behavior of failed nodes in MPI is deliberately unspecified. The same
concepts are used for shared-memory as well. Bit-vectors (32-bit or
64-bit) are split into reader and writer parts and atomic operations
are used to update the value~\cite{mellor1991scalable}.

To tolerate client failures, all lock holders have to store their
pid. A stored pid is the proof that the lock is taken. Our data
structure's design is shown in the bottom part of
\figref{fig.lockdesigns}. For a $k$-reader single-writer lock with
$f$-fairness, we use an array of $k$ 64~bit slots to hold the readers
$r_0$--$r_{k-1}$, one slot for the writer $w$, $f$ slots for the
waiting queue $\mathit{sl}_0$--$\mathit{sl}_{f-1}$, and one slot for
the outer lock ($\mathit{lock\ owner}$), which has to be acquired to
modify the lock data structure itself. Each entry is either 0 or the
pid of the respective client. Using the lock-holder's pid to indicate
whether a lock is taken allows clients to use failure detectors on
lock-holders. If a client wants to acquire a taken lock, it can either
be taken because of contention or because the holder failed. The
client starts a failure detector on the lock-holder and retries to
acquire the lock to cover both cases.

Fairness (equal share and no starvation) would require a waiting queue
with sufficient capacity to hold all waiting clients, which is a
theoretical but not a practical solution. A compromise is $f$-fairness
with a waiting queue of length $f$. Clients in the queue are subject
to fairness. They cannot overtake each other. The first node is always
the next to acquire its desired lock. However, we cannot guarantee
fairness for clients waiting to enter the queue.

When the desired lock becomes available, the process in the first slot
of the queue takes the outer lock, takes the desired lock, copies the
other members of the queue one step forward, sets the last element to
zero, and releases the outer lock. All operations require atomic CAS
operations, as a crash of the writing process during a non-atomic RDMA
write may result in a slot with a valid pid of a process not intending
to hold a lock.

Due to the 64~bit size limitation of remote atomics, it is not
possible to shift the complete queue in a single CAS operation.
Therefore, the outer lock is needed to prevent other processes from
interfering.  During the shift, each process in the queue is always
stored at least once in it---either in the old and/or new
slot. If the shifting process fails, the fairness in the queue is
preserved. On success, the last slot becomes zero and is available for
the next client. Entering the queue does not require holding the outer
lock. It is a CAS with a zero entry, which may fail. A process holding
a read or write lock can release it at any time by zeroing its
slot. It does not have to acquire the outer lock beforehand.

\section{Simulating Power-Failures}
\label{sec.testing}

To validate our implementation, we simulate power failures by killing
processes with \texttt{SIGKILL} following the approach used by
NV-Heaps~\cite{Coburn:2011:NMP:1961296.1950380}. We start a
process doing insert resp.\ remove operations. At a later time, we
kill the process. Similar to a power failure, we lose all transient
data. As the process and thus the mmapped address range does not exist
anymore, dirty cache lines cannot be written back. The process could
have been killed at any instruction of any state
transition. Afterwards, we start a new process that recovers the
current state and progresses to the clean state. The two processes
actually use the same code. While the former assumes that it is in the
clean state, the latter actually reads the current state from the
file.

The testing revealed an issue with idempotence. We killed a process
during a tree rotation. The recovery process obviously executed the
tree rotation again, but tree rotations were not idempotent at that
time. As discussed in \secref{sec.rbtrees}, failures during tree
rotations can lose sub-trees. The challenge with tree rotations is
that they read and write the same memory locations. In the example, we
could lose access to Q, the pivot, or B. To make them idempotent, we
have to store all read values in the \redolog{}. Thus, we have to keep
pointers to all three of them in the \redolog{} to separate
the read and write sets. Since then, we have run more than 2,000,000
tests without revealing any further issues.

\section{Evaluation}\label{sec:evaluation}

For all experiments with NVDIMM-N and Infiniband, we used one server
with two Intel Xeon Gold 6138 and one server with two Intel Xeon
Silver 4116 CPUs. Each server has 192\,GiB main memory and two 16\,GiB
NVDIMM-N. They are connected with an \infiniband{} FDR network
(ConnectX-3). We used CentOS 7.5, Clang~6.0.0, PMDK~1.5.1, and
UCX~1.5.

For each measurement, we report the median among 1,000 samples. The 99
percent confidence interval (CI) is always within the 1.5 percent of
the reported medians. Extremely short runs show slightly larger
percentages.

\trees{} are self-balancing binary search trees. Each node has at most
two children. There are no high radix variants. They do not amend
themselves to the optimizations commonly used for B and
B+-trees. \trees{} are simply not competitive. Our contributions are
not in the area of optimizations for speed, but we designed a new
transaction system with $\mathcal{O}(1)$ log-space, in-place updates,
and RDMA. Thus, we did not compare the performance of our
implementation with highly optimized B+-trees. We want to analyze our
optimizations and the scalability of our approach itself. We used the
data structures as shown in \figref{fig.rbtree-fsm} and
\figref{fig.avlt-fsm}. The constant-size \redologs{} were used for
trees from 0 to $10^7$ nodes.

\subsection{Local Performance for \trees{} with NVDIMM-N\label{sec:evaluation:local}}

For all local experiments, we used the Xeon Gold. As
\figref{fig.rbtree-fsm} shows, sequences through the \sm{} often
alternate between flushing to the log and reading the log back to
update the tree. This concept is needed to facilitate idempotent
updates of the tree. This style of micro-transactions performs well
with persistence mechanisms based on write-backs without invalidation
and can take full advantage of cache hierarchies. According to
\citet{pcommit}, \texttt{clwb} should show such behavior, but our
evaluation shows no significant differences between \texttt{clwb} and
\texttt{clflushopt}---neither in micro-benchmarks nor in the shown
\tree{}  benchmarks. \textsf{clflush} is consistently slower
than \textsf{clwb} and \textsf{clflushopt}.

For all local benchmarks, we varied the number of keys inserted into
empty trees. The range covers 10 to $10^6$. Keys were inserted in
random order. For remove, we re-used the filled tree and removed all
items in a different random order.

\begin{figure}[h!]
  \centering
  \includegraphics[width=\textwidth]{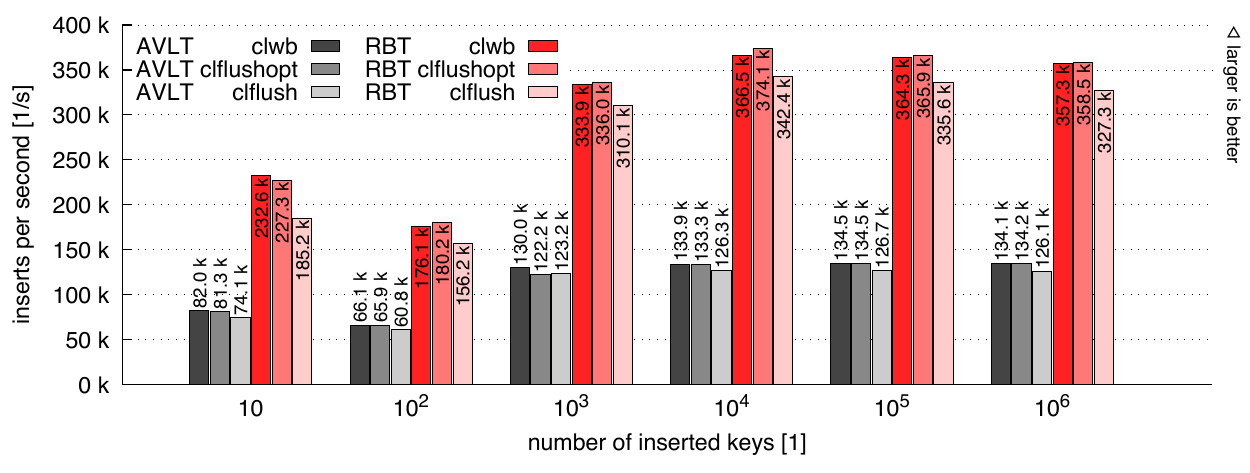}
  \caption{Insert throughput for \trees{} both with look ahead with NVDIMM-N.\label{fig.avlt-rbt-insert}}
\end{figure}

For small trees, the performance is worse than for larger trees, cmp.
\figref{fig.avlt-rbt-insert}. For trees larger than 100 keys, the
performance stabilizes. It could be due to the fact that for small
trees insert operations touch a larger share of the tree, i.e., they
flush out a large part of the tree. For larger trees, large parts of
the tree remain untouched and can be accessed without cache misses in
the next operation. If you insert a key on the left side, it will
evict parts of the tree on the path down from the caches. If the next
insert is on the right side, there will be only a low number of cache
misses. If the tree is small, the two paths will overlap and cause
more cache misses.

\begin{figure}[h!]
  \centering
  \includegraphics[width=\textwidth]{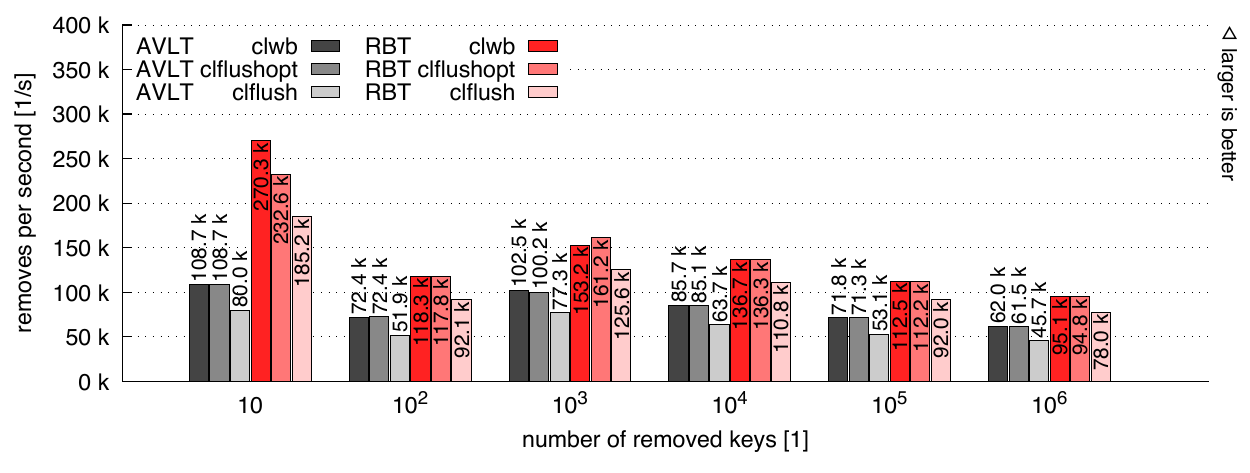}
  \caption{Remove throughput for \trees{} both without look ahead with NVDIMM-N.\label{fig.avlt-rbt-remove}}
\end{figure}

According to \citet{TARJAN1983253}, \trees{} only need
$\mathcal{O}(1)$ balance operations per insert on average. With the
optimizations described in \secref{sec.optzs}, we almost eliminate the
logarithmic part of the insert operation. As expected, the insert
costs are independent of the size of the tree. It also shows again
that \rbts{} are faster than \avlts{}. \rbts{} might reduce the costs
for insert by leaving the trees less balanced than \avlts{}, see
\secref{sec.trees}. Despite \citet{cormen2009introduction,avltrees}
suggesting that the throughput should decrease with the depth of the
tree, we can keep it constant.

The costs for remove are much higher than for insert, cmp.
\figref{fig.avlt-rbt-remove}. Note that we only tried to optimize
insert operations. The expected costs per remove are $\mathcal{O}(\log
N)$. \trees{} perform $\mathcal{O}(1)$ balance operations
per remove on average. As expected \avlts{} are slower than
\rbts{}. The gap is much smaller.

\begin{figure}[h!]
  \centering
  \includegraphics[width=\textwidth]{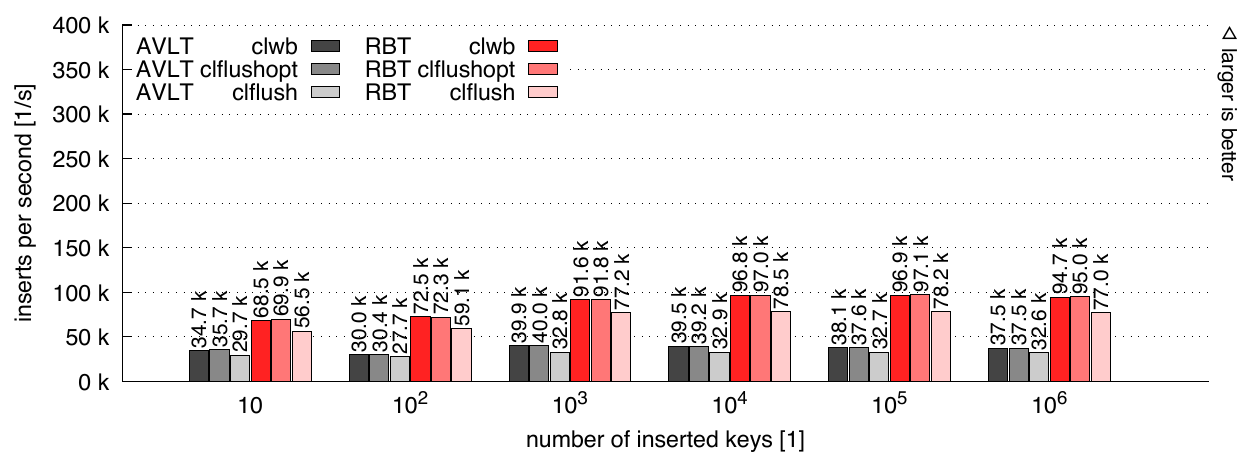}
  \caption{Insert throughput for \trees{} both with look ahead with Intel Optane.\label{fig.avlt-rbt-insert-optane}}
\end{figure}

\begin{figure}[h!]
  \centering
  \includegraphics[width=\textwidth]{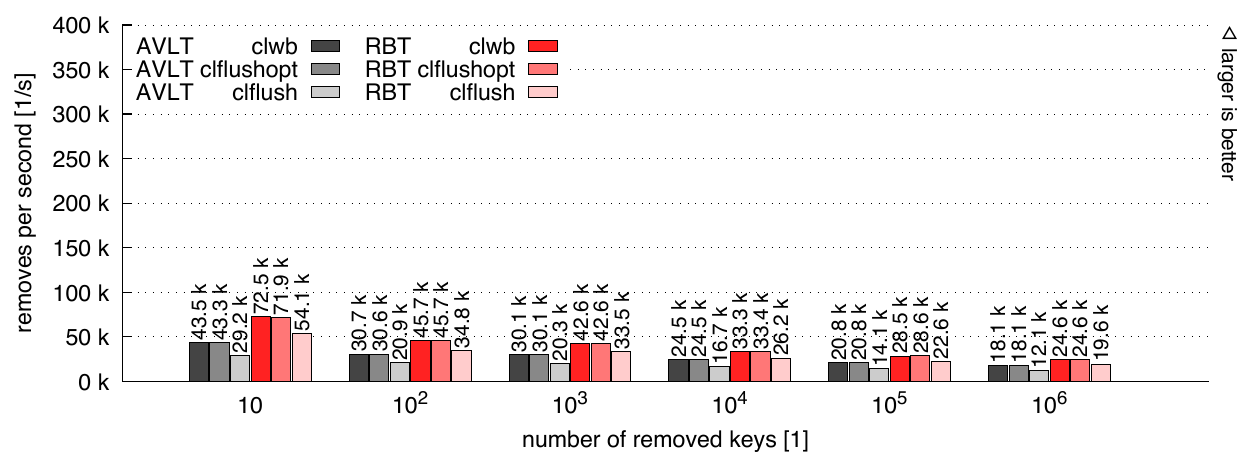}
  \caption{Remove throughput for \trees{} both without look ahead with Intel Optane.\label{fig.avlt-rbt-remove-optane}}
\end{figure}

For the evaluation of their RBT
code~\citet{Wang:2018:PRN:3190860.3177915} simulated NVDIMM, STT-RAM,
and PCM. For insert with $10^6$ resp.\ $2^{10}$ keys, they achieved:
666,666, 166,666, and 52,600 inserts per second. It shows that the
workload is latency sensitive. For simulating NVDIMMs, they used plain
DRAM. Our hardware setup differs from theirs, but our NVDIMM-Ns run at
a lower clock than DRAM. With close to 400,000 inserts per seconds for
RBTs, we are close despite a completely different approach. Our
\avlts{} are slightly slower because they keep the tree more balanced
and thus require more epochs.

\subsection{Local Performance for \trees{} with Intel Optane\label{sec:evaluation:localaep}}

For the experiments with Intel Optane, we used one server with two
Intel Xeon Platinum 8260L CPUs. It has 3~TB Apache Pass. We used
CentOS 7, GCC 9.1, and pmdk 1.7.1.

According to \cite{pcommit3,yang2020empirical} Optane has a three
times higher latency than DRAM in idle mode. Under load, the latency
and bandwidth depends on the access pattern. While the shape of the
graphs for NVDIMM-Ns and Intel Optane are similar, Intel Optane is
slower by a factor of 3-5x.

The optimizations for insert keep the performance constant despite the
trees growing. For remove the performance drops with the size as
expected. In all experiments with Intel Optane the performance of
\texttt{clwb} and \texttt{clflushopt} is indistinguishable, see
\figref{fig.avlt-rbt-insert-optane} and
\figref{fig.avlt-rbt-remove-optane}.

\subsection{Gain of Optimizations}

\begin{figure}[h!]
  \centering
  \includegraphics{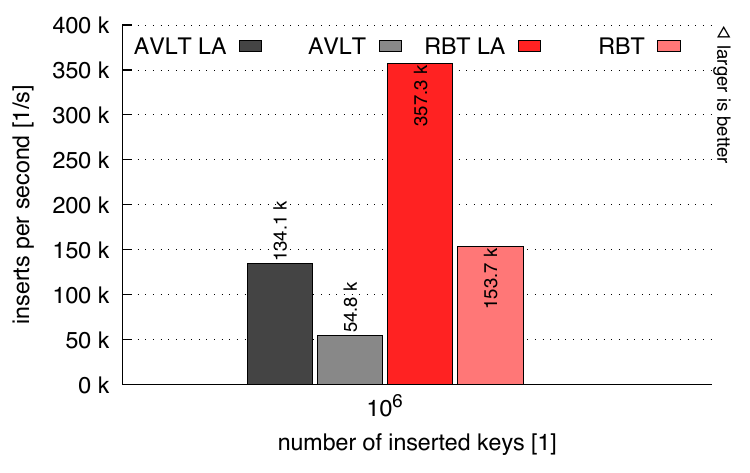}
  \caption{Insert throughput for \trees{} with and without
    look-ahead (LA) using \texttt{clwb}.\label{fig.avlt-rbt-insert-la}}
\end{figure}

To analyze the improvements of our optimizations,
see~\secref{sec.optzs}, we compared the performance with and without
look-ahead, see~\figref{fig.avlt-rbt-insert-la}. For both kinds of
trees, the gain is over a factor of 2.5x in throughput.

\subsection{Remote Performance via RDMA\label{sec.remote}}

For all experiments with \infiniband{}, the Xeon Silver served as
client, while the Xeon Gold served as server. For all measurements, we
inserted 1,000 keys in random order into an empty tree. With flushing,
every flush epoch ends with a ping pong message from the client to the
\sa{} on the server to persist the data. The workload needs
$\approx$~40,000 flush requests. With caching enabled, we cache all
read requests to the \redolog{}. The client keeps a local copy. All
write requests are applied accordingly. Accesses to the tree are never
cached.

\begin{table}[h!]
  \begin{center}
    \caption{Remote inserts via RDMA and \sa{} based flushing (RBTs).\label{tab.rdma-sa-flush}}
  \begin{tabular}{l|cc}
    \textbf{AVLT}     &  No Cache  & Cache      \\    \hline
No Flush &  2,145/s  & 2,150/s  \\
Flush    &  1,867/s  & \textbf{1,866/s}\\
\end{tabular}\quad\quad
\begin{tabular}{l|cc}
\textbf{RBT}      &  No Cache   & Cache      \\ \hline
No Flush &  2,091/s    & 2,542/s  \\
Flush    &  1,922/s    & \textbf{2,312/s}\\
\end{tabular}
  \end{center}
\end{table}

With caching and flushing enabled (\tabref{tab.rdma-sa-flush}), we
reach more than 2,300 inserts per second for RBTs and 1,800 inserts
per second for \avlts{}. Remote transaction processing is latency
bound because of gets and there is no efficient way for remote
persisting with a passive target yet. The unsatisfactory performance
of the cache is due to the fact that accesses to the tree are not
cached. Furthermore, it does not affect communication with the
software agent.

\section{Related Work\label{sec.relatedwork}}

\textbf{Trees on \nvm{}.} A number of systems were already proposed to
manage tree data structures with persistent memory. A popular tree
variant in this area are B+trees~\cite{comer-b-trees79}, which store
all values in the leaves. CDDS
B-Tree~\cite{Venkataraman:2011:CDD:1960475.1960480}, for example,
relies on 8-byte writes and a version system for inserts as long as
free slots are available in leaf nodes. Otherwise, it uses shadow
copying to split the node and to update the inner nodes. On recovery,
it uses its version system to discard all interrupted operations.
Similarly, NV-Tree~\cite{nvtree} stores all values in leaf
nodes. While the leaf nodes are stored in \nvm{}, here, the inner
nodes are stored in DRAM and can be restored after a power failure. To
further minimize the cost of flushing, entries in leaf nodes are
appended to the corresponding leaf node and remain unsorted. Full leaf
nodes are split using shadow copying. Rebalancing is not done
on-the-fly but as a separate operation that recreates the inner nodes
and makes them the current ones atomically.  The
wB+Tree~\cite{Chen:2015:PBT:2752939.2752947} also mitigates the costs
of flushing by keeping node entries unsorted to avoid entry movements
on insert. This allows inserts with only a few 8-byte writes and
shadowing. Two B+tree algorithms exploiting weak memory models and
allowing temporal inconsistencies are FAST and FAIR~\cite{210510}.
The Bztree~\cite{Arulraj:2018:BHL:3187009.3164147} is a multi-threaded
B+Tree. It relies on Persistent Multi-Word CAS (PMwCAS) and an
epoch-based garbage collection scheme.

For radix trees, WORT~\cite{201600} maintains a tree shape independent
of the insertion order. It neither needs nor supports
balancing. Inserting items on leaves or leaf paths and pointer updates
can be done with atomic 8-byte writes. For more sophisticated adaptive
radix trees, shadow copying is used.

Most research focuses on variants of B-trees with often more than two
children and optimizations for external memory. In most cases, new
keys can be added in leaf nodes with a few flushes without any
re-balancing.

\rbts{} in \nvm{} were discussed in
\citet{Wang:2018:PRN:3190860.3177915} for the first time. While they
implemented a more complex tree kind than before, they still relied on
shadow copying and a versioning system to distinguish between the copy
and the real tree. They maintain something close to a shadow tree with
some efforts to minimize overhead. Operations are performed on the
shadow tree. An atomic pointer update switches between the shadow and
the current tree. \citet{Wang:2018:PRN:3190860.3177915} also show that
update operations on RBTs are not local operations. It does not suffice
to do shadow copying on individual nodes. It requires shadow copying
of larger fractions of the tree.

All these systems handle requests in a blocking way and operations
often become visible with the last 8-byte atomic pointer update.
After a crash, it is challenging for them to identify aborted
resp.\ the last successful operation. Highly desirable properties such
as exactly once semantics, see~\secref{sec.detectable}, are hard to
achieve.  In contrast, with our \sm{} approach, we can guarantee
eventual success after accepting the command to the \redolog{} and
support exactly once semantics.

\textbf{Transactions on \nvm{}.} Systems supporting generic
transaction processing on \nvm{} based on redo logging are, for
example, SoftWrAP~\cite{giles2015softwrap} and
DudeTx~\cite{Liu:2018:DUT:3190860.3177920}.  They use a mix of
shadow-memory and redo logging where all memory accesses during a
transaction are aliased into a volatile memory region and writes are
stored in a persistent \redolog{} immediately or when all work of the
transaction is done. For DudeTx, the \redolog{} is then applied to the
actual data stored in persistent memory in a final step. With language
extensions, Mnemosyne~\cite{DBLP:conf/asplos/VolosTS11} provides
primitives for working with persistent memory. Variables can be marked
as persistent. Code regions marked as atomic will be executed with
durable transactions. It hooks into a lightweight software transaction
system to implement write-ahead redo logging.

Other systems base their transaction system on write-ahead and undo
logging~\cite{Shin:2017:HLL:3079856.3080240}. First, all store
operations are written to the undo log before the real transaction is
executed. In case of a power failure, uncompleted transactions are
rolled back. The NV-Heaps
system~\cite{Coburn:2011:NMP:1961296.1950380} provides its own heap
manager, specialized pointers, and atomic sections for persistent
memory. For transactions, it keeps a volatile read log and a
non-volatile write log. In case of an abort or power failure, it rolls
back all changes.

Here, the literature seems to be undecided between undo and redo
logging. However, undo logging suffers from more flush operations
compared to redo logging. Shadow memory is a neat way to exploit the
fact that there are two types of memories available---volatile and
non-volatile---with different performance characteristics. It provides
isolation and can leverage the benefits of caches. Our approach of
in-place updates is seldomly found in the literature.

\textbf{Logging in Databases.}  The quasi-standard algorithm for
write-ahead logging (WAL) with no-force and steal policies,
ARIES~\cite{Mohan:1992:ATR:128765.128770}, has influenced the design
of many commercial databases. It is optimized for spinning disks and
maintains an append-only log. The log contains undo and redo
records. For recovery, it goes through 3 phases: (a) analyze the log
for uncommitted and aborted transactions, (b) redo finishable
transactions, and (c) undo the remaining transactions. In our
approach, we only us a fixed-size \redolog{}. While Aries uses
write-only WAL, we read the log during epochs to facilitate
idempotence. Our analysis phase simply identifies the current state
and continues from there.

While ARIES optimizes for sequential writes,
MARS~\cite{Coburn:2013:AMT:2517349.2522724} exploits the fact, that
SSDs support high random access performance. It introduces the concept
of editable atomic writes (EAW), which are essentially \redologs{}. The
full transaction is executed in a redo record and on commit the system
applies the transaction atomically. On failure, it can simply reapply
the \redolog{}. In contrast, we write the data directly into the data
structure. The transaction becomes re-doable because of the \redolog{}
created in the previous epoch. We always split large transactions into
a sequence of micro-transactions. For \nvm{}, \redologs{} provide lower
costs. They reduce the number of flushes in contrast to undo logs. We
can also avoid complex log pruning mechanisms, because the log has a
fixed size and every operation re-uses the log of the previous
operation.

\section{Discussion of Correctness\label{sec.correct}}

\ts{remove for camera ready version}

\emph{General approach.} It is a standard technique in compiler
construction to convert code into control flow graphs with basic
blocks. For the execution of this representation we can use \sms{}. It
tracks which basic black is currently executed and which are the legal
successor blocks. State transitions correspond to the execution of
basic blocks. Large basic blocks can be split into a sequence of
smaller ones without changing the algorithm. Additionally, splitting a
basic block into two: (a) reading from the data structure and writing
to the log and (b) reading the log and updating the data structure,
ought to make both basic blocks idempotent. This is a common task for
databases with appropriate logging. The size of the basic block
resp. the number of stores influences the size of the log.

\noindent \emph{\trees{}.} In \secref{sec.testing}, we discussed our
testing approach. After a crash, we verified that we can recover and
return to the clean state. Furthermore, after the recovery we checked
whether the tree is correct, i.e., correctly balanced and colored. For
all experiments in \secref{sec:evaluation}, we inserted $k$ keys into
an empty tree and removed the same $k$ keys in a different order from
the tree. This indicates that the \sms{} are correct. Otherwise this
would yield corrupt trees. During the development of the \sm{}, we
extensively tested the \sm{}. After each insert or remove operation,
we verified that the tree is correct and the number of nodes was as
expected.

\section{Discussion of Limitations\label{sec.limits}}

For remote access, message passing might in some cases provide higher
performance than RDMA, but it is a completely orthogonal approach to
shared memory for local access. It would require two completely
different transaction systems, but our goal was to design one
transaction system for local and remote usage. As shown in
\secref{sec.impldetails}, we abstract from local and remote access and
use one common implementation for both.

For B and B+ trees, insert, remove, and balancing are operations of
limited scope. They do not need transactions. The algorithms for these
trees are of low complexity and the corresponding \sms{} would be
tiny. They are tuned for absolute performance. The literature went for
high radix trees for performance reasons with low balance
costs. However, for \trees{} balancing is the common
case~\cite{TARJAN1983253}. \trees{} are simply not competitive and
serve other demands.

As discussed before, data structures that need an auxiliary space,
which is not in $\mathcal{O}(1)$, cannot be supported with a constant
size log. Allocating additional memory would violate our
assumptions. The only option left is store the auxiliary data in the
data structure. AVL remove needs a stack of $\mathcal{O}(\log{n})$. We
use the common technique of maintaining pointers to parents. In
theory, that would induce space overhead in each tree node. The
initial data layout for the \rbts{} was 32~bytes. It had sufficient
unused space to add up-pointers (\texttt{Up} and \texttt{Dir}) without
needing to change its size, see~\figref{fig.common-tree}. Insert and
remove in trees can often be implemented with top-down algorithms,
which only need constant-sized auxiliary space. Linked lists and
hash-tables also need constant-sized auxiliary space. \ts{extend}

Atomic CAS for \nvm{}~\cite{203358,Pavlovic:2018:BAP:3212734.3212783}
brings its own challenges. Atomic operations are commonly used,
because (a) they do no tear and (b) provide protection against
concurrent access. In our approach, we are only interested in the
former property, because we use locks for thread safety. In this
paper, we consider CAS more like a read, modify, and atomic update
resp.\ store operation of 8~bytes (cmp.~\secref{sec.systemmodel}) as
there is no contention.

\section{Conclusion}

We presented a new transaction system for complex data structures in
\nvm{} that provides exactly once semantics and linearizable
durability with a \redolog{} of constant size. It splits large
transactions into smaller micro-transactions and uses a \sm{} approach
to perform larger transactions step-wise. Every accepted transaction
will eventually succeed and will never be aborted. For local and
remote access, we use the same primitives: load, store, atomic update,
and persist. This allowed us to design one transaction system that
runs locally and with \infiniband{} for remote access. As our approach
is not prepared for concurrent access, we use locks to control
concurrency. For remote access, we designed a fault-tolerant lock with
$f$-fairness.

\citet{Wang:2018:PRN:3190860.3177915} showed the first \rbt{}
implementation for \nvm{}, but their approach is based on shadowing
the whole tree.  We presented, to the best of our knowledge, the first
\avlt{} implementation for \nvm{} and the first \rbt{} implementation
for \nvm{} without shadowing and with updates `in-place'. These trees
are algorithmic far more challenging than the trees covered in the
literature so far. Insert and remove are global operations instead of
a sequence of operations with limited scope. Thus, they need
transactions.

Shadowing can claim that data structures are consistent all the
time. This approach atomically replaces parts of a data structure with
new data. Intermediates steps are not
visible. \citet{Wang:2018:PRN:3190860.3177915} atomically replace the
old tree with the new one. There is also no need for recovery, but it
fails at exactly once semantics. In our approach, the data structure
might be inconsistent after a crash, but it is recoverable all the
time. We see recovery as finishing the interrupted operation, i.e.,
moving forward to the clean state. Likewise, we support exactly once
semantics. By using a constant-sized log, we avoid any overhead for
dynamic log allocation, log pruning, and keeping a shadow copy of the
whole tree.

\section{Availability}

Our code is on
GitHub~\footnote{\url{https://github.com/tschuett/transactions-on-nvram}}
%% URL does not work.
under the Apache License 2.0.

\section*{Acknowledgments}
  The authors thank ZIB's Supercomputing department and ZIB's
  core facilities unit for providing the machines and
  infrastructure for the evaluation.  This work received funding from
  the German Research Foundation (DFG) under grant RE~1389 as part of
  the DFG priority program SPP~2037 (Scalable data management for
  future hardware).  This work is partially supported by Intel
  Corporation within the “Research Center for Many-core
  High-Performance Computing” (Intel PCC) at ZIB.

\bibliographystyle{plainnat}
\bibliography{ms}

\end{document}